\PassOptionsToPackage{pdfpagelabels=false}{hyperref}

\documentclass[useAMS,usenatbib]{mnras}

\usepackage{amssymb,natbib,graphicx,subfigure,color,ctable,times}

\def\LCDM{$\Lambda\mbox{CDM}$}

\def\AREPO{{\small AREPO}}

\def\SUBFIND{{\small SUBFIND}}
\def\SUBLINK{{\small SUBLINK}}

\def\log{{\rm\thinspace log}}

\newcommand{\ifm}[1]{\relax\ifmmode#1\else$\mathsurround=0pt #1$\fi}
\newcommand{\be}{\begin{equation}}
\newcommand{\ee}{\end{equation}}
\newcommand{\bea}{\begin{eqnarray}}
\newcommand{\eea}{\end{eqnarray}}

\newcommand{\Fig}[1]{Fig.~\ref{f:#1}}
\newcommand{\Figs}[2]{Figs.~\ref{f:#1} and \ref{f:#2}}

\newcommand{\sggg}[1]{\textcolor{green}{[]}}

\def\dex{{\rm\thinspace dex}}

\def\um{{\rm\thinspace \mu m}}
\def\nm{{\rm\thinspace nm}}

\def\pc{{\rm\thinspace pc}}
\def\kpc{{\rm\thinspace kpc}}

\def\Mpc{{\rm\thinspace Mpc}}

\def\kms{{\rm\thinspace km\thinspace s}^{-1}}

\def\Msun{\hbox{$\rm\thinspace M_{\odot}$}}

\def\yr{{\rm\thinspace yr}}

\def\Gyr{{\rm\thinspace Gyr}}

\def\Msunpc2{{\Msun\pc}^{-2}}
\def\Msunyrkpc2{{\Msun\yr^{-1}\kpc}^{-2}}

\def\magarcsec2{{\rm\thinspace mag\thinspace arcsec}^{-2}}

\title[Galaxy Size Evolution in IllustrisTNG]{The Size Evolution of Star-forming and Quenched Galaxies in the IllustrisTNG simulation}

\author[Genel, S. et al.]
{\parbox{20cm}{
Shy Genel$^{1,2}$\thanks{E-mail: shygenelastro@gmail.com},
Dylan Nelson$^{3}$,
Annalisa Pillepich$^{4}$,
Volker Springel$^{5,6}$,\\
R{\"u}diger Pakmor$^{5}$,
Rainer Weinberger$^{5}$,
Lars Hernquist$^{7},$
Jill Naiman$^{7}$,\\
Mark Vogelsberger$^{8}$,
Federico Marinacci$^{8}$,
and
Paul Torrey$^{8}$
}\vspace{0.3cm}\\
$^{1}$Center for Computational Astrophysics, Flatiron Institute, 162 Fifth Avenue, New York, NY 10010, USA\\
$^{2}$Columbia Astrophysics Laboratory, Columbia University, 550 West 120th Street, New York, NY 10027, USA\\
$^{3}$Max-Planck-Institut f{\"u}r Astrophysik, Karl-Schwarzschild-Straße 1, 85740 Garching bei M{\"u}nchen, Germany\\
$^{4}$Max-Planck-Institut f{\"u}r Astronomie, K{\"o}nigstuhl 17, 69117 Heidelberg, Germany\\
$^{5}$Heidelberg Institute for Theoretical Studies, Schloss-Wolfsbrunnenweg 35, 69118 Heidelberg, Germany\\
$^{6}$Zentrum f{\"u}r Astronomie der Universit{\"a}t Heidelberg, ARI, M{\"o}nchhofstr. 12-14, 69120 Heidelberg, Germany\\
$^{7}$Harvard-Smithsonian Center for Astrophysics, 60 Garden Street, Cambridge, MA 02138, USA\\
$^{8}$Department of Physics, Kavli Institute for Astrophysics and Space Research, MIT, Cambridge, MA 02139, USA\\
}

\begin{document}

\maketitle

\label{firstpage}

\begin{abstract}
We analyze scaling relations and evolution histories of galaxy sizes in TNG100, part of the IllustrisTNG simulation suite. Observational qualitative trends of size with stellar mass, star-formation rate and redshift are reproduced, and a quantitative comparison of projected r-band sizes at $0\lesssim z\lesssim2$ shows agreement to much better than $0.25\dex$. We follow populations of $z=0$ galaxies with a range of masses backwards in time along their main progenitor branches, distinguishing between main-sequence and quenched galaxies. Our main findings are as follows. (i) At $M_{*,z=0}\gtrsim10^{9.5}\Msun$, the evolution of the median main progenitor differs, with quenched galaxies hardly growing in median size before quenching, whereas main-sequence galaxies grow their median size continuously, thus opening a gap from the progenitors of quenched galaxies. This is partly because the main-sequence high-redshift progenitors of quenched $z=0$ galaxies are drawn from the lower end of the size distribution of the overall population of main-sequence high-redshift galaxies. (ii) Quenched galaxies with $M_{*,z=0}\gtrsim10^{9.5}\Msun$ experience a steep size growth on the size-mass plane after their quenching time, but with the exception of galaxies with $M_{*,z=0}\gtrsim10^{11}\Msun$, the size growth after quenching is small in absolute terms, such that most of the size (and mass) growth of quenched galaxies (and its variation among them) occurs while they are still on the main-sequence. After they become quenched, the size growth rate of quenched galaxies as a function of time, as opposed to versus mass, is similar to that of main-sequence galaxies. Hence, the size gap is retained down to $z=0$.
\end{abstract}



\section{Introduction}
\label{s:intro}
The size of any particular galaxy reflects various physical processes in its evolutionary history that play a fundamental role in galaxy formation. First, galaxy size is believed to be related to the angular momentum content of the galaxy \citep{MoH_98a}, which is obtained on large scales from cosmological tidal torques \citep{FallS_83a} and further affected by various dynamical processes in the non-linear regime \citep{DeFelippisD_17a}. Second, galaxy mergers affect galaxy size by scrambling stellar orbits, depositing accreted mass and triggering new star-formation. For example, dry minor mergers are believed to increase galaxy size by building an outer envelope \citep{NaabT_09a}, wet mergers to trigger a compact star-burst \citep{HernquistL_89a}, and gas-rich mergers may also contribute to further star-formation in a large post-merger disc \citep{RobertsonB_06a}. Third, gravitational instabilities inside galaxies themselves may lead to changes in size, as mass flows towards the galaxy center \citep{EfstathiouG_82a,DekelA_14b}. Conversely, the size of a galaxy can affect other aspects of its further evolution. For example, compact galaxies may be prone to quenching either by enhancing the \citet{ToomreA_64a} stability of the disc \citep{MartigM_09a}, or by harboring a particularly powerful AGN at their centers \citep{RangelC_14a}. Smaller star-forming galaxies have higher surface densities than larger ones, possibly promoting more vigorous galactic winds that affect the further activity of the galaxy \citep{MurrayN_10b}.

Galaxy size is a basic observational property that is found to significantly vary with galaxy mass, color or star-formation activity, and redshift \citep{ShenS_03a,FergusonH_04a,TrujilloI_06a,ElmegreenD_07a,WilliamsR_10a,MoslehM_12a,OnoY_13a,vanderWelA_14a,LangeR_15a}. These dependencies encode important information on the formation processes of galaxies of different types. It has generally been found that: {\bf (i)} more massive galaxies tend to be larger, {\bf (ii)} late-type/star-forming galaxies are larger than early-type/quenched ones, and {\bf (iii)} galaxies are smaller at higher redshifts, more so for high-mass than for low-mass galaxies, as well as more so for quenched than for star-forming galaxies\footnote{Though some studies, in particular early ones, found very little size evolution with time at a fixed mass \citep{LillyS_98a,BardenM_05a,IchikawaT_12a,StottJ_13a,CurtisLakeE_14a}.}. There appears to be an intimate relation between star-formation activity and size, in that most star-formation occurs within a narrow ($\lesssim0.3\dex$) size range (at a given stellar mass and redshift), and there exists a threshold in central density beyond which galaxies are quenched \citep{KauffmannG_03a,FranxM_08a,vanDokkumP_15a,BluckA_16a,WhitakerK_17a,MoslehM_17a}.

However, systematics have plagued attempts to derive a fully self-consistent observational picture. Such systematics include definitions of galaxy masses and sizes \citep{DuttonA_11a}, background subtraction \citep{HeY_13a}, light fitting procedures \citep{StottJ_11a,MoslehM_13a,BernardiM_12a}, wavelength dependence \citep{KelvinL_12a,BondN_14a,vanderWelA_14a}, selection effects \citep{LawD_12a,MoslehM_13a}, and possibly cosmic variance. 
In this work we make an attempt to minimize systematics when comparing our simulation results to observations, but we stop short of full forward modeling of the simulation to match the parameters of particular observational studies.
Regardless, it is advisable to keep in mind that systematic uncertainties due to the factors described above still play a role in the comparison between different observations as well as between observations and theoretical models.

On the theoretical side, the simplest idea is that galaxy size is proportional to halo virial radius, as a result of conservation of angular momentum during collapse and cooling \citep{MoH_98a}. Indeed, such a linear relation was inferred using abundance matching, with a proportionality ratio of $\approx0.015$ between galaxy half-mass radius and halo virial radius, over eight orders of magnitude in galaxy mass at $z=0$ \citep{KravtsovA_13a}. \citet{SomervilleR_17a} recently extended this kind of analysis up to $z\sim3$ and found proportionality ratios $\sim2-3$ times larger, with some variation with mass and redshift. \citet{HuangK_17a} reproduced the \citet{KravtsovA_13a} result for early-type galaxies (also up to $z\sim3$) but found an approximately twice larger proportionality value for late-type galaxies. A model like this may account for the scatter in galaxy sizes at a given stellar mass \citep{SomervilleR_17a} and for the stronger size dependence on redshift that is displayed by higher-mass galaxies \citep{StringerM_14a}. The non-negligible variations of the proportionality factor with mass and redshift, however, require an explanation, as does the difference between late-type and early-type galaxies (considering halo spin, for example, can at most partly account for this difference; \citealp{RomanowskyA_12a}). Such explanations need to consider the specific physical processes that control the mass and size evolution of galaxies over time.

Disc galaxies are usually considered to be growing inside-out by continuous star-formation that is fed by accretion from outside the galaxy \citep{PichonC_11a,BirdJ_13a}, and indeed the effective sizes of the star-forming parts of disc galaxies are observed to be typically larger than the effective sizes of the existing stars \citep{NelsonE_12a}. If in addition to the approximate angular momentum conservation of this accreted gas, one also takes into account the inner structure of dark matter halos (due either to `plain' \LCDM{ } halo concentrations or potentially also to adiabatic responses to baryons, which are stronger at later times), it is possible to reproduce the observed weaker time evolution of the sizes of disc galaxies as compared to the virial radii of halos \citep{SomervilleR_08b,FirmaniC_09a,DuttonA_11a}. Recent attention has however been given to the possibility that high-redshift disc galaxies may sometimes significantly shrink in size due to gravitational instabilities \citep{ZolotovA_15a}.

For early-type galaxies, the size evolution is usually interpreted using different considerations. The typical size of galaxies with any given selection criteria (and in particular, quenched galaxies in a given mass bin) will evolve with time due to two factors: i) the size evolution of individual galaxies that correspond to the selection, and ii) the transition of galaxies into and outside of the selection. A particular scenario within this context that is extensively debated in the literature is `progenitor bias', according to which the time evolution of the typical size of quenched galaxies is driven by the appearance of new, large quenched galaxies as a result of the ongoing quenching of (large) star-forming galaxies \citep{CarolloC_13a,CassataP_13a,KrogagerJ_13a,BruceV_14a}. Other studies concluded however that `progenitor bias' is sub-dominant compared to the size growth of individual galaxies \citep{HopkinsP_10c,BernardiM_11a,TrujilloI_11a,HuangS_13a}. Such individual size growth has most commonly been attributed to (in particular, minor) mergers (\citealp{NaabT_09a,OserL_11a,OogiT_13a}; but see \citealp{NipotiC_12a}), however also expansion due to mass loss has been suggested (driven by feedback as well as stellar evolution; \citealp{FanL_08a,FanL_10a,vanDokkumP_14a}).

In their toy model, \citet{vanDokkumP_15a} assume that galaxies, as a population, evolve in parallel tracks defined by $r\propto M^{0.3}$ as long as they are star-forming, even if individual galaxies may undergo periods of stronger growth or even a decrease in size. Once galaxies reach a certain (redshift-dependent) central density (or velocity dispersion), they quench, and from then on evolve on a steeper track on the size-mass plane, $r\propto M^2$. \citet{vanDokkumP_15a} found that this model can approximately explain the evolution of the observed distribution of quenched and star-forming galaxies on the size-mass plane between $z\approx3$ to $z\approx1.5$, although some tensions remain.

While analytical considerations and toy models like the ones discussed above have a paramount role in developing our understanding, our ultimate goal is to explain galaxy sizes within a full galaxy formation model in a cosmological context. Semi-analytical models based on \LCDM{ }merger trees, as well as cosmological hydrodynamical simulations, include a wide variety of physical processes relevant for galaxy formation and their non-linear interactions, and nowadays can produce semi-realistic galaxy populations. Importantly, they also allow one to directly follow the evolution of individual galaxies and their sizes over cosmic time. Studies of galaxy sizes using these models are, however, still scant.

Specifically, investigations using hydrodynamical simulations, which as opposed to semi-analytical models make no assumptions about galaxy sizes and have no parameters that control them explicitly, have considered sizes only in limited regimes. For example, they focused on a single redshift \citep{SalesL_10a,McCarthyI_12a} or a particular mass and size selection \citep{WellonsS_14a,WellonsS_16a}, or used a small number of zoomed-in halos \citep{JoungM_09a,OserL_11a,BrookC_12b}, where only the evolution of individual systems can be probed, but not statistics of populations for different cosmic epochs. Also, most have not separated galaxies of different types, and usually included minimal exploration of how and why the relations emerge. Notable exceptions to this last point are \citet{BrooksA_11a}, who found that individual simulated disc galaxies grow along the size-mass relation such that the relation hardly evolves with time; \citet{OserL_11a}, who explicitly showed the contribution of dry mergers to size growth; and \citet{DuboisY_13a}, who showed that massive galaxies are significantly larger in zoom-in simulations that include AGN feedback than in those that do not.

Recently, several studies addressed the sizes of the galaxy populations in the EAGLE cosmological simulations \citep{SchayeJ_14a}. \citet{CrainR_15a} found that galaxy sizes can serve as important constraints on feedback models, \citet{FerreroI_17a} showed that realistic sizes and their evolution are crucial for reproducing the Tully-Fisher relation and its evolution, and \citet{DesmondH_16a} found that galaxy sizes depend very weakly on halo properties, at a given stellar mass. \citet{FurlongM_17a}, separating active and quenched galaxies, found a good match between the sizes of EAGLE galaxies and observed ones. Individual star-forming galaxies were followed backwards in time and found to grow roughly on the observed relation, but the progenitors of individual quenched galaxies were found to evolve much more slowly, except for the most massive ones. Compact quenched galaxies at high-redshift disappear with time in EAGLE, growing by a combination of outward stellar migration, mergers, and subsequent star-formation.

Galaxies with $M_*\lesssim10^{11}\Msun$ in the original Illustris simulation \citep{VogelsbergerM_14a,VogelsbergerM_14b,GenelS_14a} are larger than observed galaxies by roughly a factor of $2$ \citep{SnyderG_14a,BottrellC_17a,FurlongM_17a}. Although galaxy size evolution was studied using that simulation in certain regimes \citep{WellonsS_14a,WellonsS_16a}, this offset precluded a reliable study of the size evolution of galaxies as a whole. Here we explore the size evolution of galaxies and relations between size, mass, and star-formation activity in the recently completed TNG100 simulation, which is part of the IllustrisTNG suite\footnote{www.tng-project.org} \citep{SpringelV_17a,PillepichA_17a,NelsonD_17a,NaimanJ_17a,MarinacciF_17a}. The discrepancy with observations of galaxy sizes in the original Illustris simulation is now much improved with the TNG model, an advance that comes about due to a combination of several modifications to the galactic winds model \citep{PillepichA_16a}. In this work we do not explore the origin of this improvement and its dependence on the physical models or resolution. Instead, we will use the good agreement between observations and TNG100 as a basis for a study of galaxy size evolution in this simulation as an `effective', and close to realistic, cosmological model.

The paper is organized as follows. In Section \ref{s:methods} we provide a brief description of the simulation and numerical methods used in the analysis. In Section \ref{s:comparison} we present basic size relations in the simulation and a comparison to observations. In Section \ref{s:evolution} we study evolution trends of simulated galaxies. In Section \ref{s:discussion} we discuss our results and in Section \ref{s:summary} summarize them and conclude.

\section{Methods}
\label{s:methods}
\subsection{The Simulation}
\label{s:simulations}
TNG100 \citep{SpringelV_17a,PillepichA_17a,NelsonD_17a,NaimanJ_17a,MarinacciF_17a} is built and improves upon the Illustris simulation, and uses the same $(\sim110\Mpc)^3$ cosmological box (converted to Planck cosmology) at essentially the same resolution ($\approx1.4\times10^6\Msun$ baryonic). The simulation is evolved using the \AREPO{ }code \citep{SpringelV_10a} and follows self-gravity, magnetohydrodynamics, radiative cooling, star-formation, and an updated set of sub-grid physics models for stellar evolution, black hole growth, and stellar and AGN feedback \citep{WeinbergerR_16a,PillepichA_16a}.

We developed the fiducial physical model used to evolve TNG100 with the help of a series of tuning tests on smaller cosmological boxes, with the goal of roughly reproducing several galaxy scaling relations. In this tuning process, the only aspect of galaxy size we considered was the $z=0$ overall size-mass relation, and the comparison did not take into account important factors that are discussed below in Section \ref{s:comparison} for a proper comparison between simulation and observation, and has therefore been approximate. Further, the tuning process did not involve any actual parameter value choice that was motivated by the galaxy sizes. Rather, the sizes emerged as different from the original Illustris simulation as a consequence of a combination of various new components of the model (for details, see \citealp{PillepichA_16a}). Consequently, the rich phenomenology presented here, where galaxy size is compared against mass, redshift and star-formation activity, and the evolutionary tracks of individual galaxies over time, are emergent rather than imposed.

\subsection{Analysis}
\label{s:analysis}
A simulated `galaxy' for the purposes of this study is a \SUBFIND{ }halo \citep{SpringelV_01}, each of which is tagged as a `central' or a `satellite'\footnote{Of the satellites included in the analysis, \SUBFIND{ }halos with $<10\%$ of their mass in dark matter are excluded, because such objects, which are exclusively of $M_*\lesssim10^{10}\Msun$, tend to be self-bound components of galaxies, which therefore appear as independent \SUBFIND{ }halos in spite of not being true satellite galaxies of cosmological origin.}.
Individual galaxies are followed backwards through cosmic time along their `main progenitor branch' in the `baryonic' version of the \SUBLINK{ }merger trees \citep{Rodriguez-GomezV_14a}. Some galaxies tagged as a `central' at $z=0$ may have regardless been identified as a `satellite' at an earlier time. Those are, for example, `splashback' galaxies. We generally include all these types of galaxies in our analysis except where specifically noted otherwise. In Section \ref{s:evolution}, however, we exclude galaxies that experienced a significant \SUBFIND-related `switch' where its massive envelope has been temporarily assigned to another galaxy \citep{BehrooziP_15a,Rodriguez-GomezV_14a}. We do this crudely by discarding galaxies that experience a $>0.5\dex$ stellar mass drop between two adjacent snapshots, which we have verified to indeed remove galaxies with unphysical jumps along these tracks. This removal affects the evolutionary tracks we present in any visually discernible way only at the very massive end, and in a mild way that is inconsequential for our results.

We only discuss in this work the sizes of the stellar component of galaxies rather than their gas or dark matter. For each galaxy we calculate the three-dimensional half-mass radius, $R_{\rm *,3D}$, based on the evolved (namely including burning stars and stellar remnants) masses of all the stellar particles assigned to the galaxy (i.e.~gravitationally bound to the subhalo as determined by \SUBFIND). In addition, we calculate two-dimensional half-light radii in the r-band, $R_{\rm r,2D}$, by projecting the simulation box along a random direction with respect to the orientation of each individual galaxy\footnote{In general, the projected size of a galaxy varies with viewing angle in a way that depends on its axis ratios and light profile, which in turn are correlated with galaxy type and mass (e.g.~\citealp{PriceS_17a}). We leave a detailed discussion of these trends outside the scope of this paper. Instead, we take the minimal approach that is required for a fair comparison of projected sizes with observations, namely define $R_{\rm r,2D}$ as a circular radius that contains half of the light in a random projection.}. The r-band stellar luminosities are calculated as in \citet{VogelsbergerM_13a}.

We use two definitions for the stellar mass of a galaxy, one including all the stellar particles assigned to it, which is the definition used where not noted otherwise, and one including only those within $2R_{\rm *,3D}$ of its center. The mean ratio between the former and the latter ranges between $1.25$ ($\approx0.1\dex$) for $M_*\sim10^9\Msun$ galaxies to $1.55$ ($\approx0.2\dex$) for $M_*\sim10^{12}\Msun$ galaxies. For the star-formation rate (SFR) of each galaxy, we use the sum of the instantaneous SFRs of all the gas cells assigned to it.

Throughout the paper, we characterize galaxies according to their specific SFR (sSFR), by either tagging them as `main-sequence' or `quenched' according to a prescribed cut, or by quantifying their distance from the ridge of the star-formation main-sequence, $\Delta{\rm SFMS}$. These characterizations proceed as follows.

In Section \ref{s:comparison}, the ridge of the main sequence is defined, for simplicity, as $\log({\rm sSFR [\Gyr^{-1}]})=-0.94,-0.85,-0.35,0.05,0.35$ for $z=0,0.1,1,2,3$, respectively. These values correspond to the mean sSFR of galaxies with $10^{9}\Msun<M_*<10^{10.5}\Msun$ at each of these redshifts, as described in more detail in Appendix \ref{s:main_sequence}. Galaxies are then defined to be `main-sequence' if their sSFR is within $\pm0.5\dex$ of this ridge ($|\Delta{\rm SFMS}|<0.5\dex$), which corresponds to $\approx1.5\sigma$ at $z=0$ and $\approx1\sigma$ at $z=2$, where $\sigma$ is the difference between the median and the $16^{\rm th}$ percentile of the sSFR distribution (Appendix \ref{s:main_sequence}). Galaxies are defined as `quenched' if their sSFR is at least $1\dex$ below that ridge ($\Delta{\rm SFMS}<-1\dex$).

In Section \ref{s:evolution}, the distance of each galaxy from the main-sequence at its $z\geq0$ redshift is calculated with respect to a more elaborate, mass-dependent definition of the ridge. First, all galaxies within a stellar mass bin of width $0.1\dex$ around the galaxy in question are identified, and among these all that have a zero SFR are discarded. Then, galaxies in the $10\%$ tails of the remaining sSFR distribution are discarded as well. The mean sSFR of the remaining galaxies is defined as the zero-point $\Delta{\rm SFMS}=0$, namely serves as the reference main-sequence ridge the galaxy in question is compared with. Galaxies with $M_*>10^{10.5}\Msun$ are treated differently in that the zero-point $\Delta{\rm SFMS}=0$ used for them has the same sSFR value as the zero-point for $M_*=10^{10.5}\Msun$ galaxies, since the main sequence ceases to exist around that mass.

\section{Results: Size-Mass Relations and Comparison to Observations}
\label{s:comparison}
\subsection{Size-Mass Relations in TNG100}
\label{s:relations}
\begin{figure}
\centering
\includegraphics[width=0.475\textwidth]{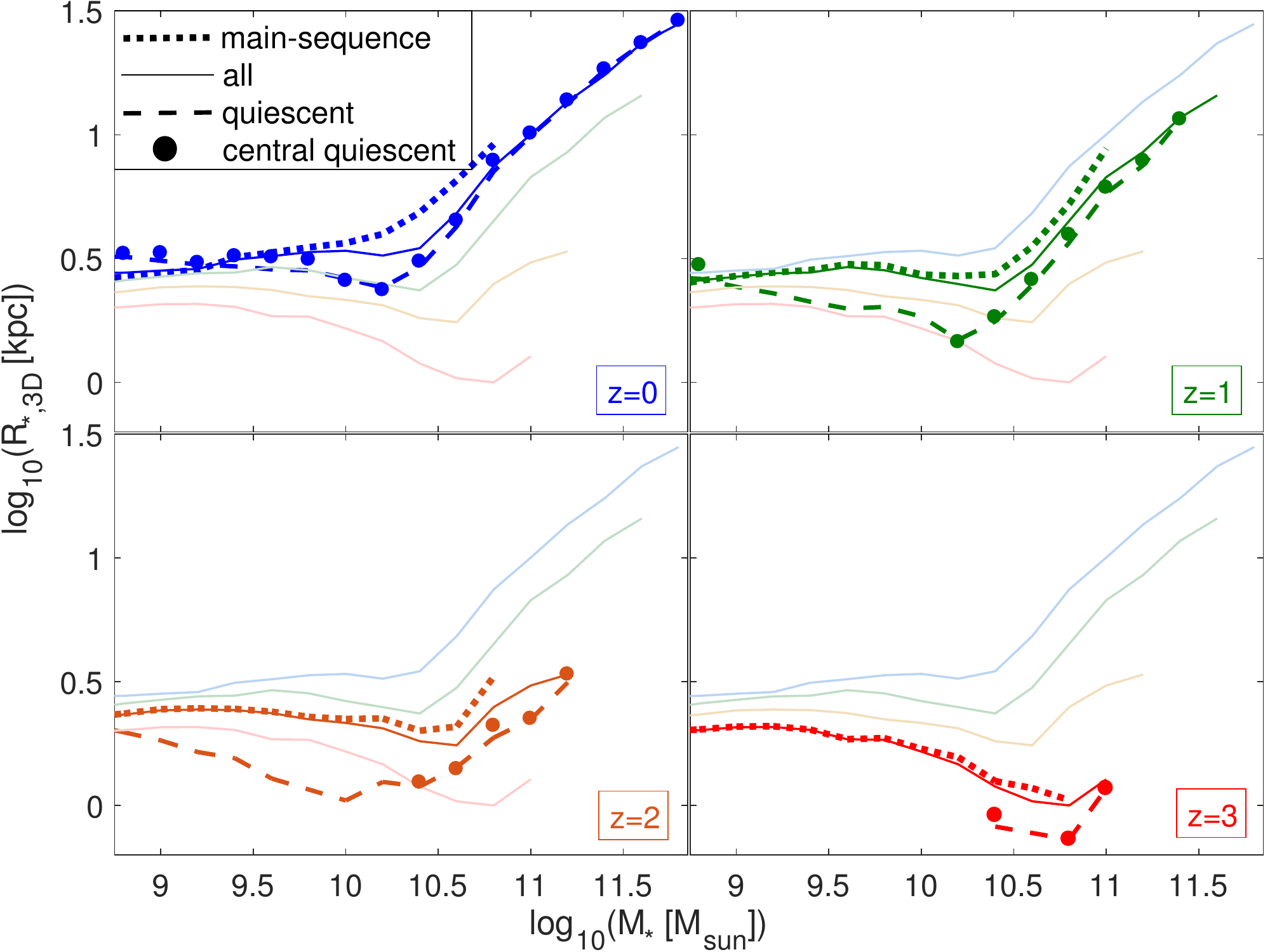}
\caption{Median size-mass relations in TNG100 based on three-dimensional half-mass sizes at redshifts of $z=0,1,2,3$. Quenched galaxies (long-dashed), in the median, are always smaller than main-sequence galaxies (short-dashed), except at $z=0$ at $M_*\sim10^9\Msun$. The difference is most pronounced, $\sim0.2\dex$, at intermediate masses, around $M_*\sim10^{10}\Msun$. When quenched {\it central} galaxies are considered alone (circles), their median size matches that of the total quenched population very well in the regime of overlap, namely above a certain (redshift-dependent) mass limit where central quenched galaxies exist in the simulation. Data is shown only for bins containing more than ten galaxies.}
\vspace{0.3cm}
\label{f:size_mass_z0_to_z3}
\end{figure}

We begin by presenting the raw simulation results in the form of median size-mass relations based on three-dimensional half-mass sizes in \Fig{size_mass_z0_to_z3}. Each panel focuses on one of four redshifts, and for each of them four galaxy selections are considered, as indicated in the legend. Solid curves, which represent the relations for the full galaxy population at each redshift, are repeated in all panels for reference. Here we use the mass definition that is limited to within $2R_{\rm *,3D}$ in order to minimize mass bias between centrals and satellites due to the inclusion/exclusion of an extended stellar halo.

\Fig{size_mass_z0_to_z3} exhibits the following trends.
{\bf (i)} The relations are rather flat at $M_*\lesssim10^{10.5}\Msun$, with logarithmic slopes of up to $\pm0.1$ that become gradually more negative toward higher redshifts.
{\bf (ii)} At $M_*\gtrsim10^{10.5}\Msun$ the relations steepen up to logarithmic slopes of $\approx0.7$.
{\bf (iii)} At any given mass and for all galaxy selections, galaxies at higher redshifts are smaller than at lower redshifts.
{\bf (iv)} At all redshifts and masses (except $M_*\lesssim10^{9.2}\Msun$ at $z=0$) main-sequence galaxies are larger (in the median) than quenched ones, and this difference is largest around $M_*\sim10^{10}\Msun$.
{\bf (v)} There is little size difference between central and satellite quenched galaxies, as exemplified by the relations shown for central quenched galaxies (circles) and the total quenched population (long dashed)\footnote{Satellite quenched galaxies are not shown explicitly for visual clarity but their relations show the same degree of differences as between quenched centrals and the total quenched populations.}. Central quenched galaxies, however, do not cover the full range in stellar mass: at $z\gtrsim1$ central quenched galaxies exist only at high masses, above where the slope of the relation breaks from flat to steep. Hence, the increasingly negative slope shown by $M_*\lesssim10^{10.5}\Msun$ quenched galaxies at higher redshifts represents a satellite population.
{\bf (vi)} Main-sequence galaxies at $z=3$ behave differently from those at lower redshift in that they exhibit an increasingly negative slope towards higher masses that does not bend upwards even at the highest masses. These trends are discussed in relation to observations in the remainder of this section, and in relation to median evolutionary tracks of individual galaxies in Section \ref{s:evolution}.

\subsection{Comparison of Size-Mass Relations to Observations}
\label{s:relations_comparison}
\subsubsection{Caveats}
\label{s:comparison_caveats}
Before proceeding to a comparison of the simulated size-mass relations with observations, we consider the plethora of systematic factors that have the potential to bias such a comparison. In order to make a reasonably fair comparison, we first make sure to use the same definition of galaxy size from the simulation as in the observations to which we compare: circularized projected half-light sizes, $R_{\rm r,2D}$. In particular, we compare the sizes that contain half the optical light in the r-band rather than half the mass; that are two-dimensional (namely, based on projections along a random line of sight); and are circularized (namely, averaged between the extents along different directions on the observed plane).

There are, however, further factors for which we do not make an exact even handed comparison. {\bf (i)} In terms of sample selection, we separate galaxies into `main-sequence' and `quenched' using the (redshift-dependent) sSFR limits described in Section \ref{s:analysis}, while the observations we compare to are based on either a selection in color-color space, or morphological selections. We estimate this uncertainty to account for potential systematics of $\sim0.1\dex$ in size, as this is roughly the difference between observational studies making these different selections, as well as roughly the maximum variation of size we obtain from the simulation by applying various reasonable sSFR cuts (see below). {\bf (ii)} In terms of the definition of the `total' galaxy light used for calculating its half-light size, we use the total bound light assigned by \SUBFIND, which for central galaxies can include light all the way to the virial radius, while observationally the total light is based on integration to large distances of profiles fitted to the high surface brightness inner parts. The accuracy of this comparison depends on the ability of the observed profile fits to capture correctly the actual full stellar light out to large distances. It is possible that observations underestimate the true, total bound light as predicted from the simulations, particularly compared with single-Sersic fits of close-to-exponential inner density profiles, which drop sharply and might miss an extended low surface brightness component. {\bf (iii)} In terms of emission and absorption sources, measurements in different bands can result in different sizes, as they weight the stellar populations differently. For the simulation we use the age and metallicity of the stellar particles to produce mock rest-frame r-band luminosity, and do not consider the effects of dust on the apparent size of galaxies. For the measurements, there exist systematic differences depending on the exact band that is used, and we apply empirically-derived corrections to convert them uniformly to the rest-frame r-band \citep{KelvinL_12a,SzomoruD_13a,vanderWelA_14a}, as described below. {\bf (iv)} In terms of the stellar mass assigned to each galaxy, we use the bound mass assigned by \SUBFIND, while observationally it is determined using the aforementioned profile fits to the light profiles and a conversion to stellar mass using model mass-to-light (M/L) ratios. Fortunately, the profile fitting component of this systematic uncertainty on the mass is quite degenerate with the profile fitting systematic uncertainty on the size, since the fitted profile would under/over-estimate the mass and the size in concert, leaving the size-mass relation minimally changed \citep{BernardiM_12a}. However, the uncertainty related to M/L ratios is still significant and hence stellar mass systematics may be present in our comparison probably up to $\approx0.5\dex$ \citep{ConroyC_09b,ConroyC_10a,BernardiM_17a}, which we consider to be the main source of systematic uncertainty. In addition to the systematic uncertainty associated with the mass, also a statistical error is expected as an unavoidable aspect of stellar population modeling, the possible effects of which we demonstrate quantitatively below.

\begin{figure*}
\centering
\subfigure[Main-sequence / late-type galaxies]{
          \label{f:size_mass_z0_obs_comparison_MS}
          \includegraphics[width=0.49\textwidth]{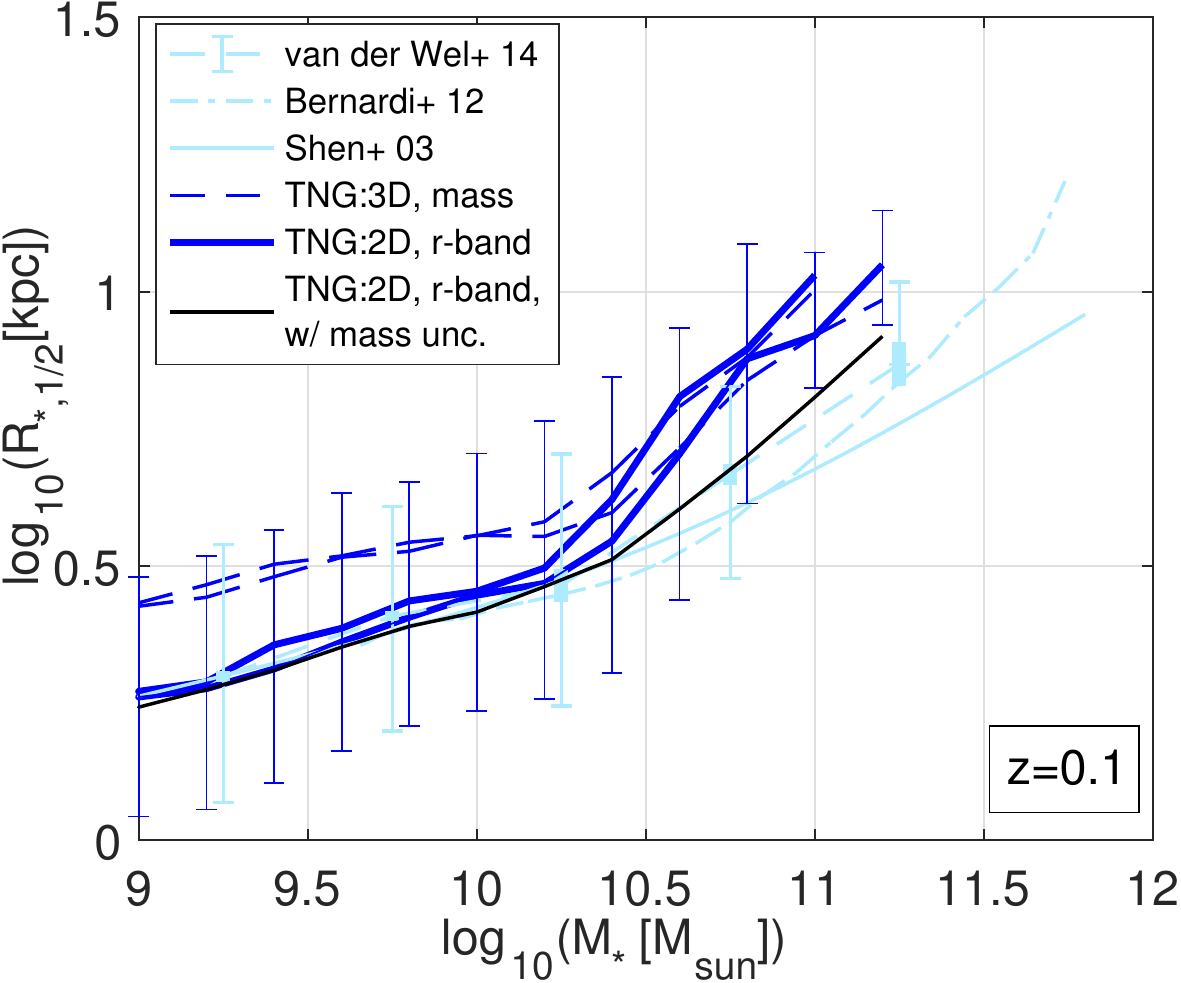}}
\subfigure[Quenched / early-type galaxies]{
          \label{f:size_mass_z0_obs_comparison_Q}
          \includegraphics[width=0.49\textwidth]{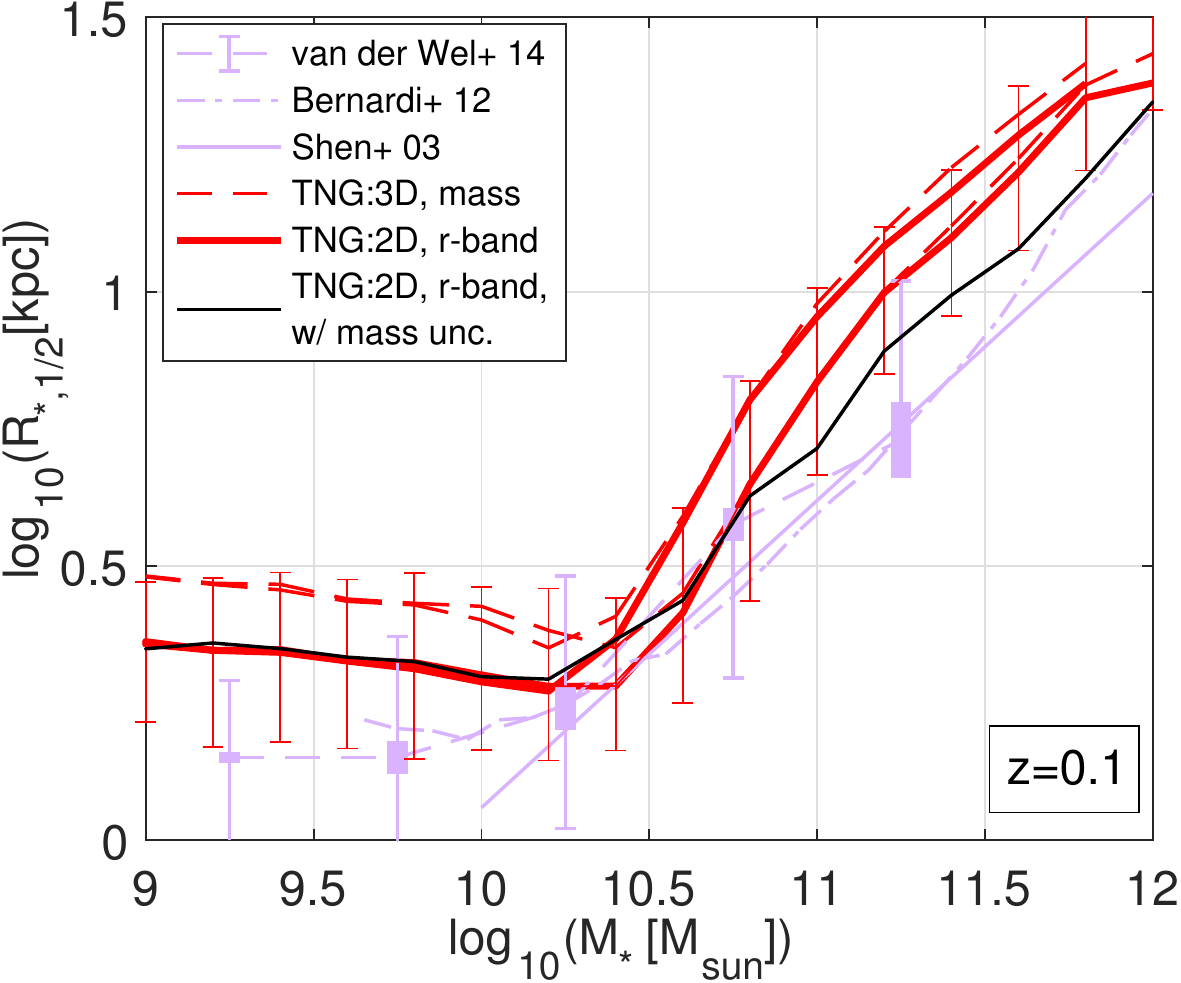}}
\caption{Median size-mass relations at $z=0.1$ compared between observations and TNG100. The distinction between galaxy types is made in TNG100 based on sSFR, while in \citet{vanderWelA_14a} on UVJ colors, and in \citet{BernardiM_12a} and \citet{ShenS_03a} on morphology. For the \citet{vanderWelA_14a} data, the thick lines represent the error on the median, and the error bars the $16^{\rm th}-84^{\rm th}$ percentiles. Only one of the simulation curves includes error bars (for visual clarity), similarly for $16^{\rm th}-84^{\rm th}$ percentiles. The primary simulation data to be considered here are the projected sizes in the r-band (solid), as this size definition most closely matches the observed one. In addition, stellar mass-based three-dimensional simulated sizes are shown for reference (dashed). For each of these two size definitions, relations using two mass definitions are shown: full mass and mass within $2R_{\rm *,3D}$ (colored curves). In addition, a relation using r-band projected sizes with the full mass (as in the right-most thick curves) is shown in concert with the effect of mock $0.25\dex$ observational uncertainty on the mass (thin black). Simulation data are shown only for bins containing more than ten galaxies.}
\vspace{0.3cm}
\label{f:size_mass_z0_obs_comparison}
\end{figure*}

\subsubsection{Observational Data Sets for Comparison}
\label{s:comparison_obs}
We consider three observed size-mass relations from the literature for comparison
with our simulation results. The main differences between these observational studies are as follows. {\bf (i)} \citet{ShenS_03a} and \citet{vanderWelA_14a} use single S\'ersic fits, while \citet{BernardiM_12a} use double S\'ersic fits that account better for extended envelopes. {\bf (ii)} \citet{BernardiM_12a} and \citet{vanderWelA_14a} perform two-dimensional fits to the whole galaxy image, while \citet{ShenS_03a} perform one-dimensional fits on azimuthally averaged annuli. This is shown by \citet{BernardiM_12a} to be responsible for their steeper slopes at high masses with respect to \citet{ShenS_03a}. {\bf (iii)} The \citet{ShenS_03a} and \citet{BernardiM_12a} measurements are based on $z\sim0.1$ galaxies from the SDSS survey, while the ones from \citet{vanderWelA_14a} are based on the CANDELS survey at $z\sim0.25$, which we extrapolate to $z=0.1$ using $R\propto H(z)^{\beta_H}$ with the mass-dependent $\beta_H$ values provided in their Table $2$, where $H(z)$ is the Hubble rate\footnote{The size-mass relation from the Galaxy And Mass Assembly (GAMA) survey reported by \citet{BaldryI_12a} based on semi-major axis sizes agrees to within a few percent with the corresponding one from \citet{vanderWelA_14a} once the latter is extrapolated to $z=0.1$ using their derived dependence on cosmic epoch. Since only \citet{vanderWelA_14a} report circularized sizes, we only use their (extrapolated) data for comparison with the simulated data, rather than the direct $z=0.1$ data of \citet{BaldryI_12a}. Due to the excellent agreement between \citet{BaldryI_12a} and \citet{vanderWelA_14a} on non-circularized data, however, we expect that if \citet{BaldryI_12a} had reported circularized sizes, they would be essentially indistinguishable in \Fig{size_mass_z0_obs_comparison} from the extrapolated \citet{vanderWelA_14a} data that is shown. Similar considerations guide us with regards to a possible comparison to \citet{LangeR_15a}.}. {\bf (iv)} The \citet{ShenS_03a} and \citet{BernardiM_12a} measurements are made directly in the SDSS r-band, while the ones from \citet{vanderWelA_14a} are made in the Hubble Space Telescope Wide Field Camera 3 (WFC3) F125W filter, centered on an observed-frame wavelength of $1.25\um$. \citet{vanderWelA_14a} report these sizes after their conversion to rest-frame $500\nm$ using the gradient $\Delta\log R/\Delta\log\lambda$ they directly measure between the F814W ($814\nm$) and F125W filters, and we further convert them using their Equation $1$ back to a somewhat longer wavelength, namely to the center of the r-band at rest-frame $623\nm$, in order to match the SDSS-based studies.
{\bf (v)} Most probably the principal reason for the systematic difference between \citet{ShenS_03a} and \citet{BernardiM_12a} versus \citet{vanderWelA_14a}, besides possible systematics on the stellar mass, is that \citet{ShenS_03a} and \citet{BernardiM_12a} make morphological selections into `late-type' and `early-type' galaxies (which are not identical but give very similar results; \citealp{BernardiM_12a}), whereas \citet{vanderWelA_14a} separate the full population based on UVJ color-color selection into `star-forming' and `passive' galaxies.

\subsubsection{Comparison Discussion}
\label{s:comparison_discussion}
In \Fig{size_mass_z0_obs_comparison} we compare these observational data to the $z=0.1$ median $R_{\rm r,2D}$-mass relation in TNG100 (thick solid). To illustrate the simulation-side aspect of the systematic uncertainty on mass (see \citet{PillepichA_17a} for an extended discussion), we present curves for both mass definitions, namely for the full bound mass, which we consider the more appropriate comparison to the observational data, as well as for the mass enclosed within $2R_{\rm *,3D}$ (smaller mass values). To demonstrate the possible effects of statistical uncertainties on observed masses, we also show the size-mass relation that results after adding a random Gaussian component to the simulated masses with a width of $0.25\dex$ (thin solid black). We find that the agreement with observations is good overall, with nominal size differences in the range $\approx0-0.2\dex$, which are generally within the uncertainties. Several regimes are noteworthy. {\bf (i)} The relations for both main-sequence and quenched galaxies have a break in the simulation at $M_*\approx10^{10.5}\Msun$, while the observations, which also show an increasing slope at higher masses, indicate a more gradual trend. However, once reasonable statistical errors in the mass measurements are considered (black), the break in the simulated relations becomes significantly less sharp and is in better agreement with the observations. {\bf (ii)} The agreement for main-sequence galaxies at $M_*\lesssim10^{10.5}\Msun$ is remarkable, but at larger masses the simulation produces galaxies that appear somewhat larger than observed. This discrepancy is again very significantly reduced once statistical errors on mass measurement are considered. {\bf (iii)} Both observations and TNG100 show an almost entirely flat size-mass relation for quenched galaxies with $M_*\lesssim10^{10.5}\Msun$, however the simulated galaxies are $\approx0.1\dex$ larger. {\bf (iv)} The agreement is excellent for quenched galaxies around a stellar mass of $10^{10.5}\Msun$, but the most massive galaxies, with $M_*\gtrsim10^{11}\Msun$, are somewhat larger in TNG100 than observed.

Beyond the median $R_{\rm r,2D}$-mass relations, two additional aspects of the simulation data are presented in \Fig{size_mass_z0_obs_comparison}. First, the relations based on the full bound mass include also error bars, which represent the $16^{\rm th}-84^{\rm th}$ percentile widths of the size distributions at a given stellar mass. The simulated widths are compared on the figure to the observed widths provided as a function of mass by \citet{vanderWelA_14a}, who however estimate that the intrinsic widths are $\sim50\%$ smaller than the observed ones for late-type galaxies and smaller by a factor of $\sim2$ for early-type galaxies. Given these corrections, the simulated widths appear to be somewhat larger than the observationally-inferred intrinsic widths. Nevertheless, they are $\lesssim0.2\dex$, which compares favourably to the robust (model-independent) conclusion quoted by \citet{vanderWelA_14a}. In addition, we find that the distribution widths of early-type galaxies are smaller than those of late-types, as inferred observationally as well. We caution, however, that the widths may be more sensitive than the medians to the population selection. We do not consider these widths further in this work, but conclude with this basic comparison that they show a reasonable agreement. Second, simulated relations based on the intrinsic size $R_{\rm *,3D}$ are shown (dashed) for comparison to the $R_{\rm r,2D}$-mass relations. At $M_*\gtrsim10^{10.5}\Msun$ they are quite similar, but this is fortuitous, as two effects roughly cancel out. On one hand, projected sizes are smaller than three-dimensional ones, and on the other hand, r-band sizes are larger than mass-based ones. For $M_*\lesssim10^{10.5}\Msun$, however, there exist more significant differences, at the level of $\approx0.2\dex$, stemming from smaller mass-versus-light size differences, which do not cancel out with the significant projection effect on the size measurement. Evidently, a proper comparison to observations must take these simple considerations into account.

\begin{figure}
\centering
\subfigure[Main-sequence, $z=1$]{
          \label{f:size_mass_z1_obs_comparison_4x4_MS}
          \includegraphics[width=0.23\textwidth]{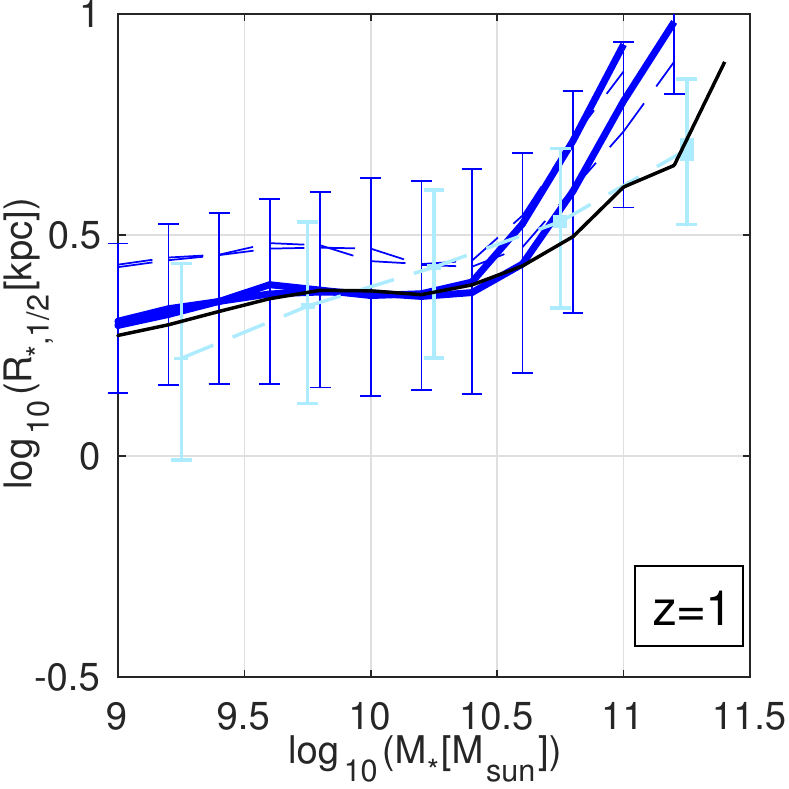}}
\subfigure[Quenched, $z=1$]{
          \label{f:size_mass_z1_obs_comparison_4x4_Q}
          \includegraphics[width=0.23\textwidth]{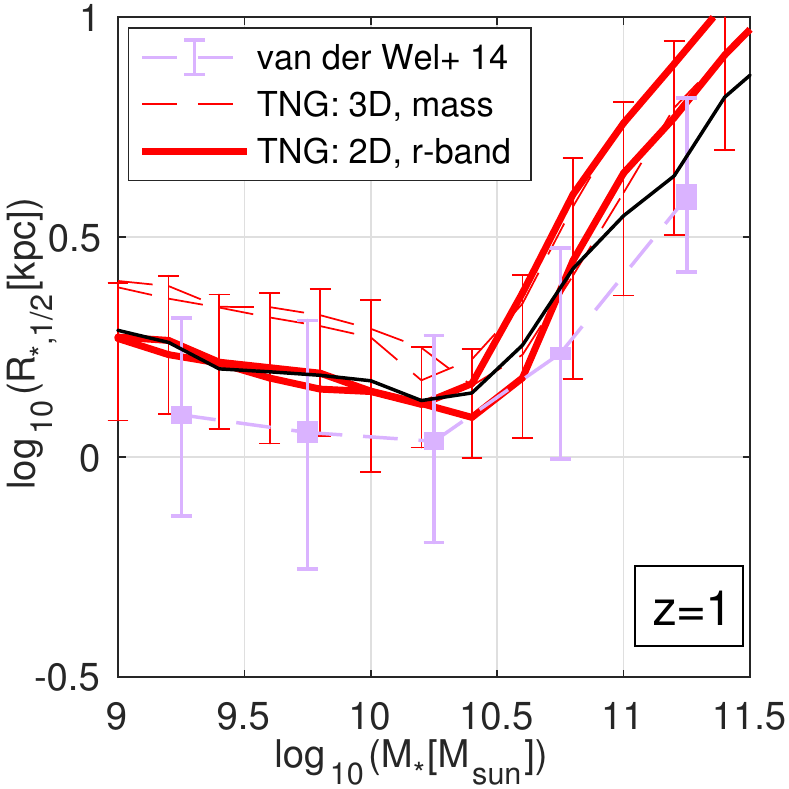}}
\subfigure[Main-sequence, $z=2$]{
          \label{f:size_mass_z2_obs_comparison_4x4_MS}
          \includegraphics[width=0.23\textwidth]{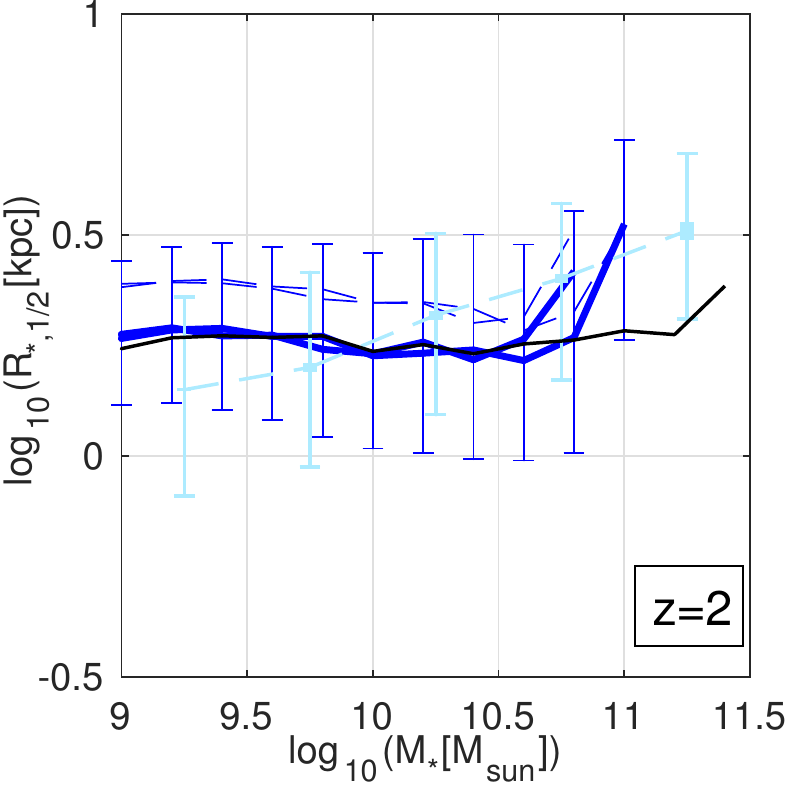}}
\subfigure[Quenched, $z=2$]{
          \label{f:size_mass_z2_obs_comparison_4x4_Q}
          \includegraphics[width=0.23\textwidth]{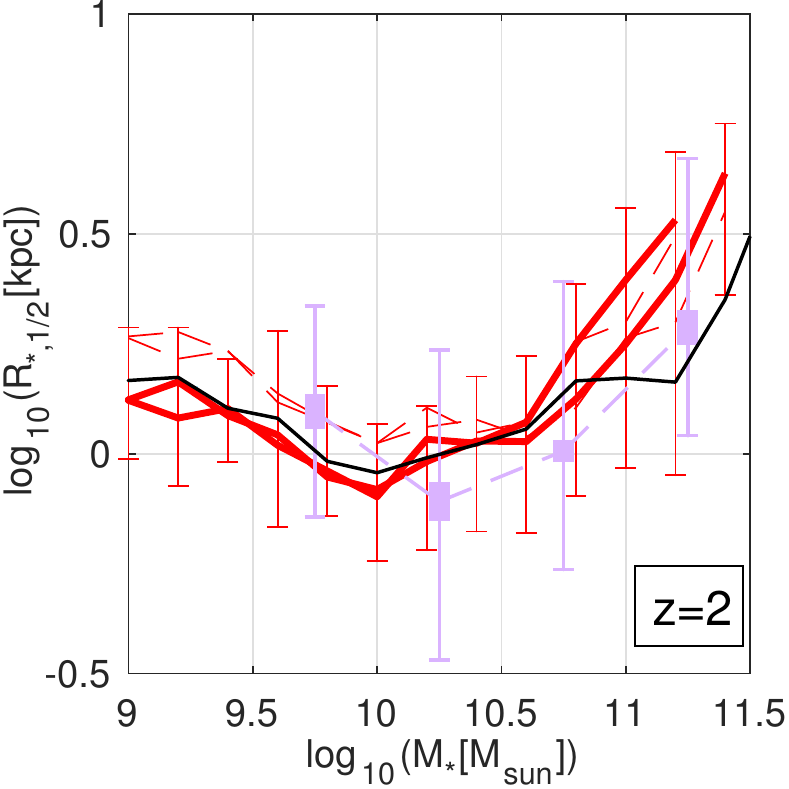}}
\caption{Median size-mass relations at $z=1$ (top) and $z=2$ (bottom) compared between observations \citep{vanderWelA_14a} and TNG100, separately for star-forming galaxies (left) and quenched ones (right). Line styles are as in \Fig{size_mass_z0_obs_comparison}, as indicated in the upper-right panel.}
\vspace{0.3cm}
\label{f:size_mass_z1_2_obs_comparison_4x4}
\end{figure}

In \Fig{size_mass_z1_2_obs_comparison_4x4} we compare median size-mass relations at $z=1$ and $z=2$ between TNG100 and the observations of \citet{vanderWelA_14a}. To derive the latter, we average (in log-space) between the nearest redshift bins reported there. The degree of quantitative agreement and general trends at $z=1$ are very similar to those at $z=0$, for both galaxy types, with the exception of low-mass main-sequence galaxies, where the simulated relation is considerably flatter than observed. This is even more so the case at $z=2$, where the simulated relation for main-sequence galaxies is even slightly negative, while the observed one has a positive slope of ${\rm d}\log R/{\rm d}\log M\approx0.15$. It is worth noting though that also in the observations themselves this $z\sim2$ slope is somewhat shallower than at lower redshifts. For quenched galaxies, the shape of the simulated and observed $z=2$ relations is essentially identical, with the simulated one shifted towards higher {\it masses} by $\approx0.2\dex$.

The overall conclusion from \Figs{size_mass_z0_obs_comparison}{size_mass_z1_2_obs_comparison_4x4} is that TNG100 reproduces the qualitative observed trends between size and mass and redshift, with appropriate differences between late-type and early-type galaxies as reflected in their star-formation activity levels. In particular, quenched galaxies are smaller than main-sequence ones, galaxy sizes become larger with cosmic time, and the slope of the size-mass relation steepens with increasing mass, while it is flat or even negative for low-mass quenched galaxies. The quantitative agreement is also good, with average differences in size of $\approx0.1\dex$ and maximum ones of $\approx0.25\dex$, which is mostly within the estimated systematic uncertainties. The spatial resolution of the simulation may play a role in some of these offsets, as Fig.~A2 in \citet{PillepichA_17a} shows that the overall $z=0$ size-mass relation with the TNG model is not fully converged, in particular at low masses, where we show here that high-redshift main-sequence galaxies as well as low-redshift quenched galaxies are larger in TNG100 than observed. There are also notable differences in the slope of the relations, where low-mass main-sequence galaxies have a relation that is too flat at high redshift (interestingly, also seen in the EAGLE simulation; \citealp{FurlongM_17a}), while at low redshift, high-mass galaxies of both types have relations that are too steep.

\begin{figure*}
\centering
\includegraphics[width=1.0\textwidth]{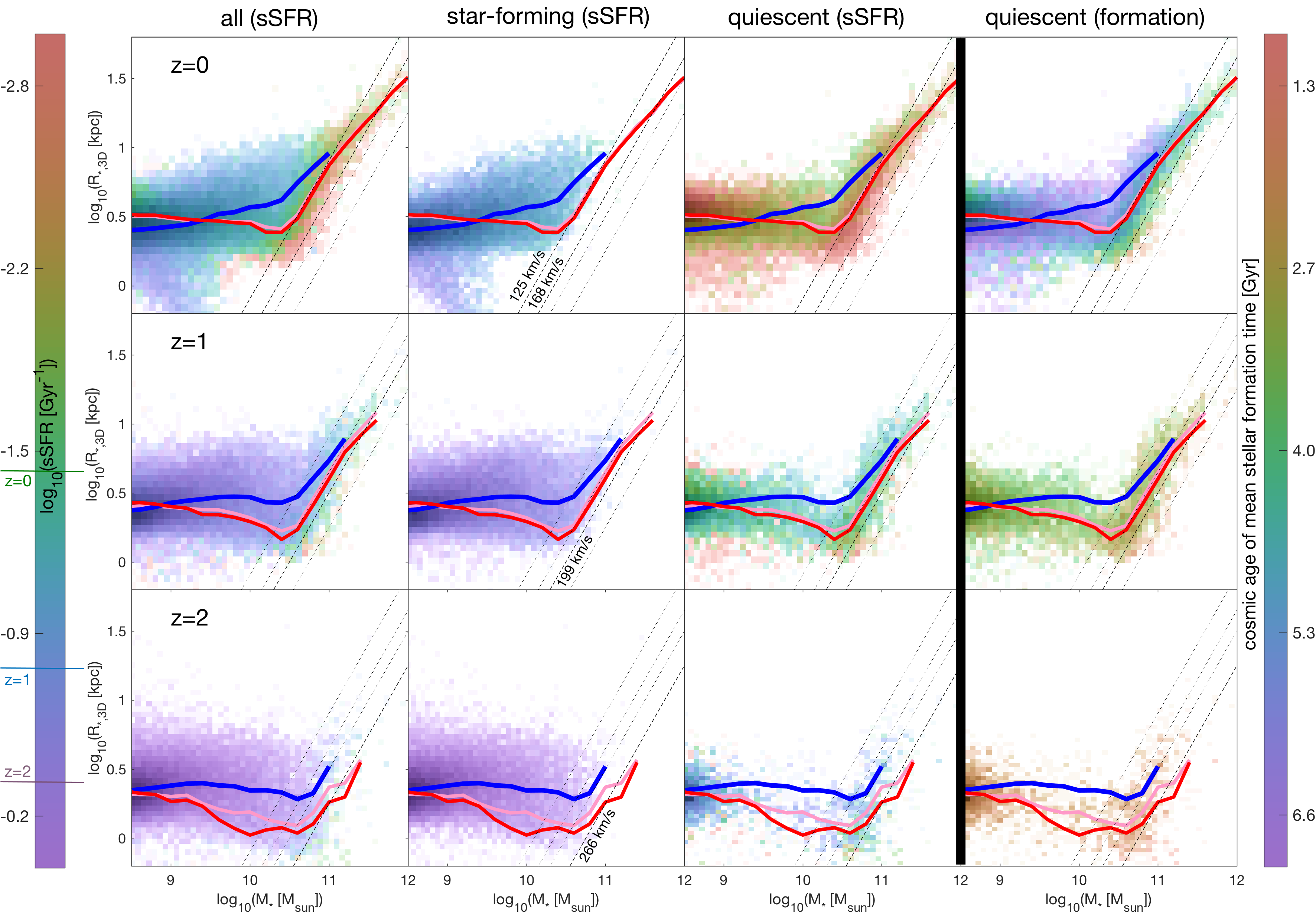}
\caption{The size-mass plane in TNG100 for three different redshifts ($z=0,1,2$ from top to bottom), color-coded by the instantaneous sSFR (three left columns) or the age of the universe at the mean stellar formation time (right column). Where all galaxies are included (left), a strong trend is seen for smaller galaxies to be less star-forming. When two sub-populations are considered separately, this trend is absent for star-forming galaxies (second from left) and weaker for quenched galaxies (third from left). Quenched $z=0$ galaxies show however a strong trend for smaller galaxies to have earlier stellar formation times (right). For $M_*\gtrsim10^{10}\Msun$, contours of constant sSFR run parallel to diagonal lines of constant gravitational potential (dashed), and the transition between star-forming and quenched galaxies occurs at higher values thereof at higher redshifts, as indicated in the second column from the left. Overlaid are median size-mass relations for star-forming (blue) and quenched galaxies (red/pink, see main text), the dividing line between which is indicated on the left color bar.}
\vspace{0.3cm}
\label{f:matrix}
\end{figure*}

\subsection{Size-Mass-SFR}
\label{s:SizeMassSFR}
\Fig{matrix} goes into more detail on relations between size, mass and star-formation in the simulated data (which includes both centrals and satellites, as in the previous two figures). For three redshifts ($z=0,1,2$, from top to bottom), the first three columns from the left show the size-mass plane where colors represent the mean sSFR in each position and brightness scales with galaxy number density on the plane. In the left-most column all galaxies are included, while the next two columns break the full galaxy populations into star-forming and quiescent galaxies, with the cuts on sSFR (indicated on the color bar on the left) applied at $0.5\dex$ below the `main-sequence ridges' defined above, namely at $\Delta{\rm SFMS}=-0.5\dex$. Note that this selection differs from the `main-sequence' versus `quenched' galaxy selection used so far, as here the combination of the second and third columns includes {\it all} galaxies. Overlaid are size-mass relations using $R_{\rm *,3D}$ and the full stellar mass. The blue and red curves are identical to the curves with the same size and mass definitions in \Figs{size_mass_z0_obs_comparison}{size_mass_z1_2_obs_comparison_4x4} (except replacing $z=0.1$ with $z=0$), while the pink curves use the same definitions, but a different selection, including all galaxies with $\Delta{\rm SFMS}<-0.5\dex$ rather than with $\Delta{\rm SFMS}<-1\dex$, as does the red curve. The differences between the red and pink curves give a sense for the sensitivity of the results to selection effects\footnote{For star-forming galaxies, the exact choice of the sSFR threshold is less important than for the quenched galaxies. Whether including only `main-sequence' galaxies ($|\Delta{\rm SFMS}|<0.5\dex$), or all galaxies with $\Delta{\rm SFMS}>-0.5\dex$ makes no visible difference in \Fig{matrix}.}. In the fourth column, the same galaxy population is shown as in the third column, but with color representing the mean stellar formation time (relative to the Big Bang) rather than the instantaneous sSFR.

\Fig{matrix} is inspired by Fig.~1 in \citet{WhitakerK_17a} and shows very similar trends to the observations from the 3D-HST survey presented there. The full galaxy population (left column) shows a strong and continuous trend of decreasing sSFR with decreasing size, at a given stellar mass in the range $10^{10-11}\Msun$. However, within each of the sub-populations of quenched and (in particular) star-forming galaxies (third and second columns from the left, respectively), this trend is weaker (see also \citealp{BrennanR_17a}). This implies that the trend for the full population is significantly driven by the changing proportions of these two sub-populations as a function of size at a given stellar mass: at smaller sizes, the proportion of quenched galaxies is larger. Quenched galaxies, in particular at low redshift, show a trend between their size and mean stellar age, such that smaller galaxies are older. This trend too is in agreement with observations \citep{WilliamsC_16a}, as well as its weakening towards higher redshifts \citep{WhitakerK_12b}. In contrast, \citet{FurlongM_17a} found no such trends of galaxy size, neither with sSFR nor with mean stellar age, for massive quenched galaxies in the EAGLE simulation. However, they did find that smaller quenched galaxies in their simulation have assembled their mass earlier, a trend we reproduce and discuss further in Section \ref{s:evolution}.

Overlaid on each panel in \Fig{matrix} are diagonal lines that represent constant values of the gravitational potential. The velocity values indicated next to these lines are based on the empirical relation describing galaxy velocity dispersion as a function of mass and size \citep{vanDokkumP_15a}. The value corresponding to the lower line, $266\kms$, is adopted from \citet{vanDokkumP_15a} as the value where the quenched fraction is about $50\%$ at $1.5<z<3$. For lower redshifts, the values are scaled following \citet{FranxM_08a}, as galaxies at lower redshifts are found observationally to quench at lower velocity dispersions (or densities; see also \citealp{WooJ_15a,WhitakerK_17a}). This scaling leads to a value of $199\kms$ at $z=1$ and $168\kms$ at $z=0$. The latter differs to a notable degree from the value where the quenched fraction is $50\%$ as directly found for $z\sim0$ galaxies by \citet{BluckA_16a} at $125\kms$, which is also marked on \Fig{matrix}.

The simulation reproduces these observations well: contours of constant sSFR correspond roughly to constant values of velocity dispersion. Moreover, the distributions of simulated galaxies shift with redshift with a trend similar to the observed one, namely the transition to quiescence (signified by the rightmost diagonal envelope of star-forming galaxies seen in the second column, as well as by the locus of quiescent galaxies seen in the third column) moves towards lower velocity dispersions with cosmic time. In the TNG model, quenching is caused by black hole feedback in the low accretion state mode \citep{WeinbergerR_16a,WeinbergerR_17a,NelsonD_17a}, the onset of which is a prescribed function of Eddington ratio and black hole mass, and hence, through the (emergent) relation between black hole mass and velocity dispersion (`the $M-\sigma$ relation'), is expected to be related to the galaxy velocity dispersion. This may provide an explanation for the trends we see here. Namely, smaller galaxies at a given mass, which tend to have higher velocity dispersions and hence more massive black holes (which indeed is the case in the simulation, though not shown here explicitly), tend to become more quenched as these black holes impart more efficient feedback. We will return to this point in the context of galaxy evolutionary histories in Section \ref{s:evolution_interplay}.

It is worth noting that the normalization of the dividing line between star-forming and quenched galaxies appears somewhat offset between TNG100 and 3D-HST, in that the simulation is shifted to lower velocity dispersions with respect to the observations by $\approx0.1\dex$. This may be explained in part by the possibility that black hole masses in TNG100 may be too large \citep{WeinbergerR_16a,PillepichA_16a}, hence the $M-\sigma$ relation in TNG100 is probably offset towards lower velocity dispersion compared to the observed relation. There could also possibly exist an offset of the $\sigma(R,M)$ relation between TNG100 and observations, a topic that is of interest for future analysis.

\begin{figure*}
\centering
\includegraphics[width=1.0\textwidth]{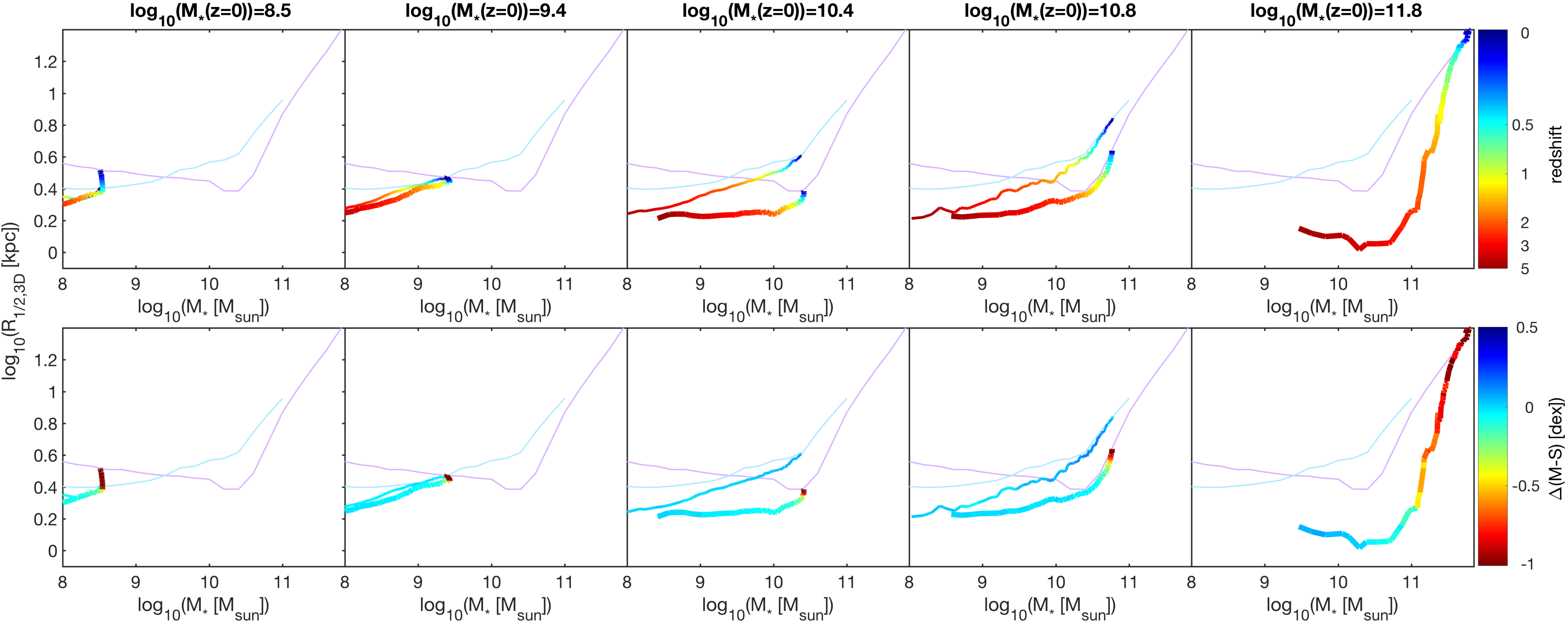}
\caption{Evolution on the size-mass plane along main-progenitor tracks of galaxy populations that contain tens to hundreds of galaxies each, and are selected with increasingly larger $z=0$ masses (left to right). All values on these tracks are medians of the respective quantities, and they are color-coded by median redshift (top) or distance from the (mass- and time-dependent) star-formation main-sequence (bottom). In each panel there appears a track for galaxies that at $z=0$ are quenched (thick) as well as for those that are still on the main-sequence at that time (thin; with the exception of the right-most panel that represents a mass where main-sequence galaxies do not exist). In addition, light magenta and cyan curves show the median size-mass relations for $z=0$ galaxies, repeated from \Fig{matrix}.}
\vspace{0.3cm}
\label{f:evolution_MSQ_sizemass}
\end{figure*}

\section{Results: Evolutionary Tracks}
\label{s:evolution}
The goal of this section is to illuminate some of the trends seen in `snapshot' relations between size, mass, and sSFR that were discussed in the previous section. This is done by following the evolution of these quantities through time in individual galaxies as they form. To focus on the emergence of the $z=0$ relations, we select galaxies at $z=0$ and follow their `main progenitor' branches in the merger trees up to $z=5$. While there undoubtedly exists useful information in the histories of individual galaxies (e.g.~\citealp{ZolotovA_15a}), we reserve such studies to future work, and here make use of the large statistical samples available to us in the TNG100 simulation with the specific goal of studying the {\it median} formation histories of galaxies in several galaxy samples that contain $\sim10-1000$ galaxies each. These samples are selected by: (i) $z=0$ star-formation activity, namely using a distinction between quenched and main-sequence galaxies, (ii) $z=0$ stellar mass, in particular in five bins around $\log(M_{*,z=0}[\Msun])=8.5,9.4,9.8,10.8,11.8$, and in some cases (iii) $z=0$ size, in particular from the upper/lower quartiles of the size distribution of some parent sample that is selected by sSFR and mass. The width of the mass bins and the number of galaxies included in each sample is provided in Table \ref{t:numgals}.

\begin{table}
\caption{Galaxy numbers and satellite fractions for the evolutionary tracks presented in \Figs{evolution_MSQ_sizemass}{evolution_MSQ_quantities}. The second column for each galaxy type gives the fraction of corresponding galaxies that are satellites at $z=0$, while the third column provides the fraction of galaxies that have {\it ever} been satellites in their history.}
\label{t:numgals}
\begin{tabular}{|c||c|c|c||c|c|c|}
\hline
& \multicolumn{3}{c||}{Quenched} & \multicolumn{3}{c|}{Main-sequence} \\
\cline{2-7}
$\log_{10}$ & $\#$ & $f_{\rm sat}$ & $f_{\rm sat}$ & $\#$ & $f_{\rm sat}$ & $f_{\rm sat}$ \\
$(M_{*,z=0}[\Msun])$ & gal. & $z=0$ & ever & gal. & $z=0$ & ever \\
\hline
8.45 - 8.55 & 541 & 86\% & 97\% & 730 & 23\% & 35\% \\
9.3 - 9.45 & 462 & 91\% & 99\% & 1493 & 27\% & 38\% \\
10.3 - 10.45 & 348 & 57\% & 63\% & 473 & 24\% & 35\% \\
10.7 - 10.85 & 340 & 40\% & 53\% & 75 & 24\% & 43\% \\
11.65 - 11.95 & 26 & 23\% & 54\% & - & - & - \\
\hline
\end{tabular}
\vspace{0.3cm}
\end{table}

\subsection{Evolutionary Tracks of Main-Sequence versus Quenched Galaxies}
\label{s:evolution_MSQ}
\Fig{evolution_MSQ_sizemass} contrasts the median evolution histories of quenched and main-sequence galaxy populations on the size-mass plane, in various $z=0$ mass bins. The highest-mass point of each evolutionary track lies by construction on the size-mass relation for either the main-sequence or the quenched $z=0$ population, which are both shown as light cyan and magenta curves, respectively. The mass bins were selected in order to explore five distinct regimes, as follows. $M_{*,z=0}\sim10^{8.5}\Msun$ represents the mass scale where quenched galaxies are larger than main-sequence ones. $M_{*,z=0}\sim10^{9.4}\Msun$ is where the relations cross, and $M_{*,z=0}\sim10^{10.4}\Msun$ is where the distance between them is maximal. $M_{*,z=0}\sim10^{10.8}\Msun$ represents the largest masses where main-sequence galaxies exist, and $M_{*,z=0}\sim10^{11.8}\Msun$ the largest masses where galaxies exist altogether in the simulation. The colors of the tracks indicate the median value of a third quantity along them, redshift in the top row and $\Delta{\rm SFMS}$ in the bottom row. Note that $\Delta{\rm SFMS}$ is calculated with respect to the `local' value of the star-formation main-sequence ridge, namely the value corresponding to the evolving median redshift and mass along each of these tracks. \Fig{evolution_MSQ_sizemass} includes both central and satellite galaxies, as we find that they show nearly identical results when examined separately. It should be noted, however, that in the two bins with the lowest mass, almost all quenched galaxies are satellites, or have been so in the past, hence there exists no relevant population of truly central quenched galaxies in the simulation in that mass range. Table \ref{t:numgals} provides the satellite fractions in the different samples.

\begin{figure*}
\centering
\includegraphics[width=0.829\textwidth]{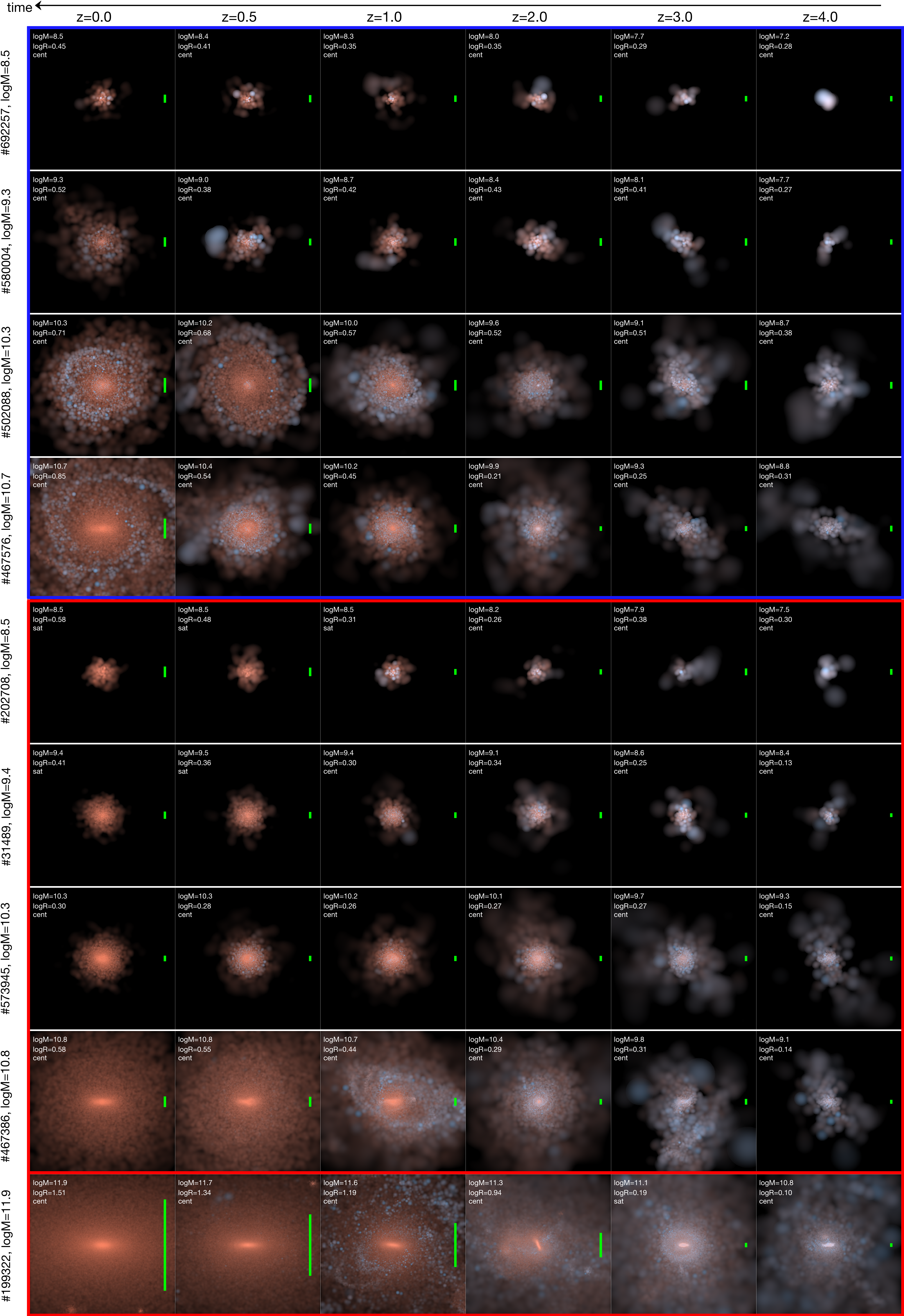}
\caption{Mock stellar light images, each $50\kpc$ on a side, of nine $z=0$ galaxies and their main progenitors at $z=0.5,1,2,3,4$, shown `face-on', i.e.~along the angular momentum vector calculated as in \citet{GenelS_14b}. The top four rows are main-sequence galaxies throughout their evolution, and the bottom five are quenched at $z=0$. The mass bins, increasing from top to bottom for each galaxy type, correspond to the ones in \Fig{evolution_MSQ_sizemass}, and match, in order, between the main-sequence rows and the top four quenched rows. The bottom row is of high mass where only quenched galaxies exist. The green bar in each panel indicates the stellar half-mass radius $R_{\rm *,3D}$.}
\vspace{0.3cm}
\label{f:images}
\end{figure*}

Before discussing \Fig{evolution_MSQ_sizemass} in depth, we point the reader to \Fig{images}, which shows mock stellar images along the evolutionary tracks of nine individual galaxies, one for each galaxy type and mass bin as in \Fig{evolution_MSQ_sizemass}. These particular galaxies were selected by eye from a parent population of ten random galaxies in each bin, such that their evolution in the size-mass-$\Delta{\rm SFMS}$-redshift space most closely matches the median trends seen in \Fig{evolution_MSQ_sizemass}. Hence \Fig{images} can serve as a visual aid to interpreting \Fig{evolution_MSQ_sizemass}.

\begin{figure*}
\centering
\includegraphics[width=1.0\textwidth]{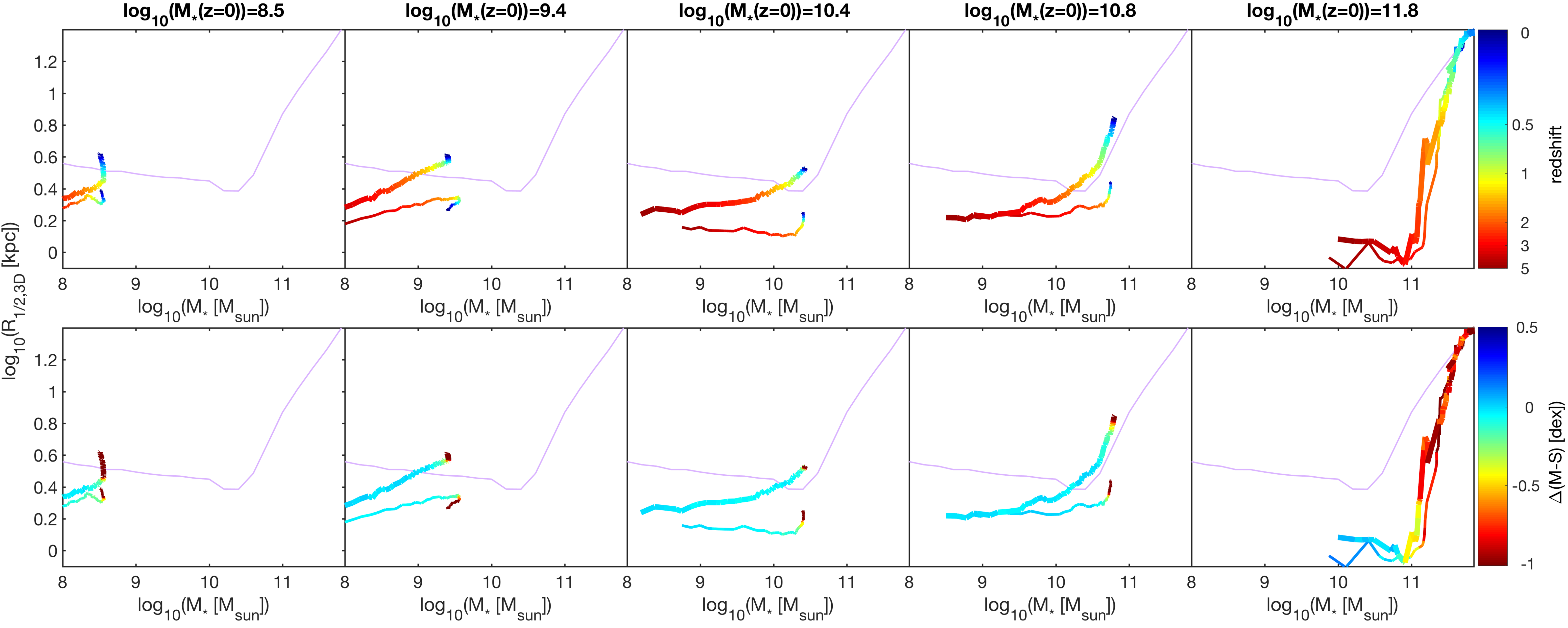}
\caption{Similar to \Fig{evolution_MSQ_sizemass}, except that only quenched galaxies are included, and the two tracks in each panel represent sub-populations thereof: the galaxies whose $z=0$ sizes are in the upper quartile of the size distribution of quenched $z=0$ galaxies (thick), and those whose sizes are in the lower quartile (thin).}
\vspace{0.3cm}
\label{f:evolution_Q_sizemass_smalllarge}
\end{figure*}

\begin{figure*}
\centering
\includegraphics[width=1.0\textwidth]{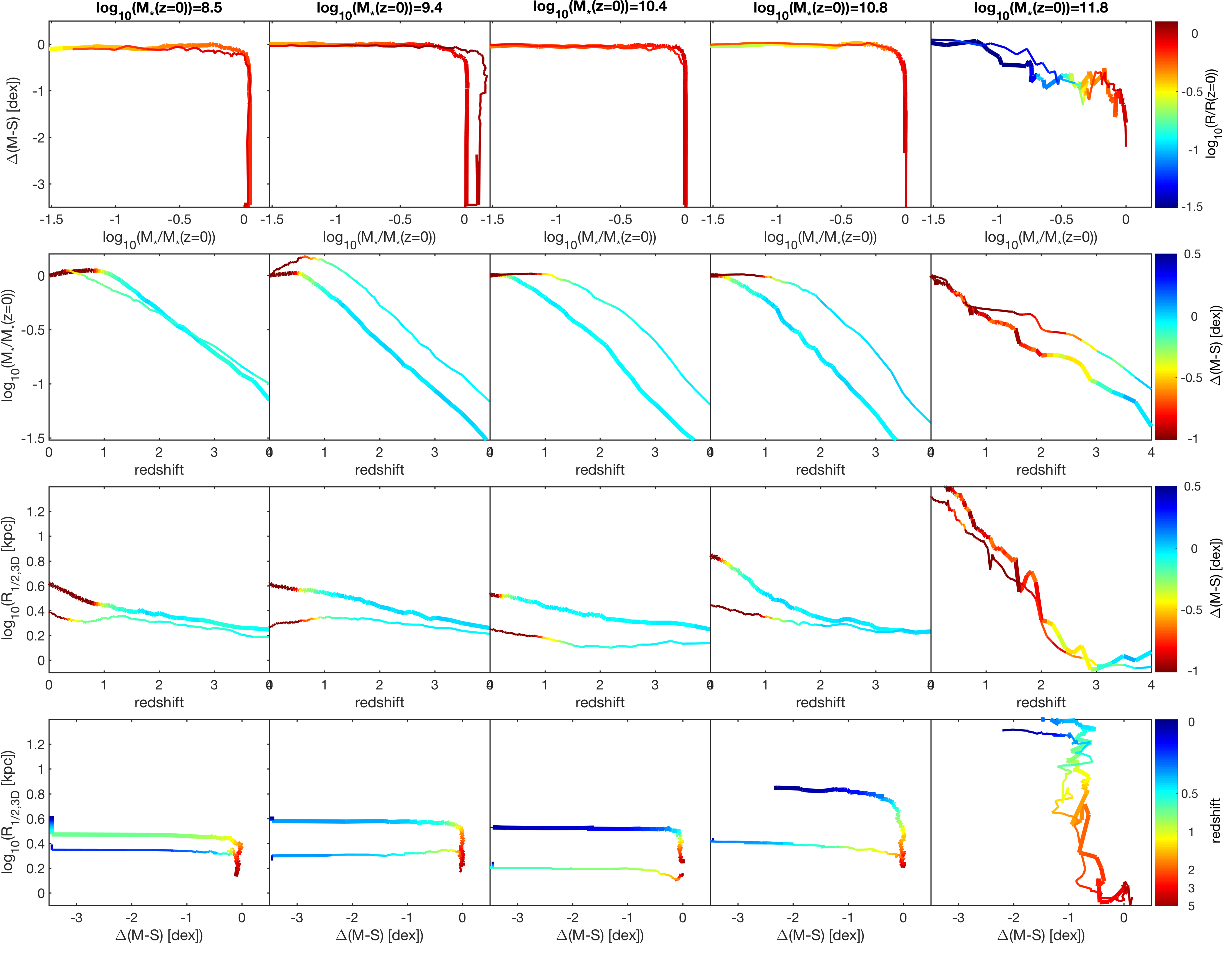}
\caption{Similar to \Fig{evolution_MSQ_quantities}, except that only quenched galaxies are included, and the two tracks in each panel represent sub-populations thereof: the galaxies whose $z=0$ sizes are in the upper quartile of the size distribution of quenched $z=0$ galaxies (thick), and those whose sizes are in the lower quartile (thin). It is seen that smaller quenched galaxies, compared to larger ones, form their mass, as well as quench, earlier in time the farthest their final mass is from $\sim10^{9}\Msun$ (third and fourth rows), and indeed they all quenched very close to their final mass, with the exception of very massive systems that quench already at about a tenth of their final mass (second row). They do not, however, simply form in the same way but at earlier time, as they reach their final size at a smaller fraction of their final mass (first row). Except for the most massive galaxies, most of the size growth, for both the larger and the smaller quenched galaxies, occurs while they are on the main sequence (third and fourth rows).}
\vspace{0.3cm}
\label{f:evolution_Q_quantities_smalllarge}
\end{figure*}

The shapes of the evolutionary tracks shown in \Fig{evolution_MSQ_sizemass} clearly depend strongly on both mass and $z=0$ star-formation activity, and generally do not coincide with the $z=0$ `snapshot' relations. The most striking feature is that the evolutionary tracks of main-sequence galaxies on the size-mass plane have similar shapes across all final mass values, while the past behaviour of $z=0$ quenched galaxies is strongly mass-dependent, and for the most part differs from that of main-sequence galaxies. The main-sequence galaxy populations have been evolving (collectively) approximately following $R\propto M^{0.15}$ for several orders of magnitude in mass and for at least as long as $10\Gyr$. Notably, their main progenitors have always been -- in the median -- essentially right on the ridge of the star-formation main-sequence, as can be read from the color of the corresponding tracks in the bottom row.

Quenched galaxies, in contrast, have not always been quenched, as may indeed be expected, and the shapes of their tracks vary strongly with mass. \Fig{evolution_MSQ_sizemass} shows that quenched galaxies in most mass bins have evolution histories that are composed of two distinct phases, separated by the quenching time itself. At low masses, $M_{*,z=0}\lesssim10^{9.4}\Msun$, the first phase looks like that of main-sequence galaxies, and the second phase involves a modest degree of size growth associated with a minor degree of mass {\it loss}. This second phase essentially goes away around $M_{*,z=0}\sim10^{9.4}\Msun$, where the size-mass tracks are almost indistinguishable between the two galaxy types, except a certain timing offset visible through the colors in the top row, which indicate the quenched galaxies formed earlier. At higher masses, the first phase, when the main progenitors of the $z=0$ quenched galaxies {\it were still on the main-sequence themselves}, is distinguished by nearly flat evolutionary tracks, namely a nearly constant size as the mass grows by over an order of magnitude. The second phase, which occurs around when $\Delta{\rm SFMS}$ drops below zero, is characterized by a sharp steepening of the evolutionary track. It however occurs at a large enough fraction of the final mass such that even with the steeper slope, quenched galaxies do not close the size gap with respect to main-sequence galaxies, a gap that was opened when the progenitors of both types were still main-sequence galaxies. Indeed only at very high masses, $M_{*,z=0}\gtrsim10^{11}\Msun$, where main-sequence galaxies do not exist anymore, do galaxies quench at a small enough fraction of their final mass (see also \citealp{NelsonD_17a}) to experience a dramatic size growth in this second, quenched, steep phase of evolution.

The evolutionary tracks in \Fig{evolution_MSQ_sizemass} show that for a given final $z=0$ mass and a given progenitor mass, the progenitors of quenched galaxies are smaller than those of main-sequence galaxies, and they also exist at an earlier time (compare track colors in the top row), namely quenched galaxies form earlier. This description holds also in the `parallel tracks' evolution scenario \citep{vanDokkumP_15a}, where the progenitors of quenched galaxies are smaller at a given mass simply because they reach that mass at an earlier time, when the overall galaxy population is smaller. The evolutionary tracks in TNG100 are, however, not parallel, as those of main-sequence galaxies have a positive slope and those of quenched galaxies are instead flat. We will return to the role of the earlier formation time of quenched galaxies in their size evolution in more detail in Section \ref{s:evolution_interplay}.

Some of these trends can be seen more explicitly in \Fig{evolution_MSQ_quantities} in Appendix \ref{s:evolution_MSQ_appendix}, which presents tracks for the same galaxy samples, but with various combinations of quantities on the horizontal, vertical, and `color' axes.

\subsection{A Focus on the Evolutionary Tracks of Quenched Galaxies}
\label{s:evolution_Q}
In \Fig{evolution_Q_sizemass_smalllarge} we focus on quenched galaxies, and present evolutionary tracks on the size-mass plane, similarly to \Fig{evolution_MSQ_sizemass}, where one track corresponds to the median evolution of the $25\%$ of $z=0$ quenched galaxies with the largest $z=0$ sizes (thick curves), and the other to the lowest quartile of quenched galaxies in terms of their $z=0$ sizes (thin curves). At $M_{*,z=0}\sim10^{8.5}\Msun$, the evolutionary tracks are similar until the quenching time, and what appears to determine the final size is the degree of late-time ($z\lesssim0.5$ in the median) size growth that occurs past quenching. At all higher masses (expect possibly the highest mass bin, where our statistical power is limited, as there are only six galaxies in each quartile), the situation is different. The evolutionary tracks are separated already at high redshift and in particular at a time when the galaxies are still on the main-sequence. Quenched galaxies that end up with larger $z=0$ sizes have steeper size-mass evolutionary tracks than those that end up smaller. This holds during the time they are still main-sequence galaxies and almost until they reach their final mass. Interestingly, at intermediate masses, the quenched galaxies that end up smaller actually have {\it steeper} size-mass evolution slopes {\it after} their quenching time. In fact, after they quenched they appear to not grow in mass at all, but still grow in size by $\approx0.15\dex$. For completeness, a figure is included in Appendix \ref{s:evolution_MS_smalllarge} that is similar but which separates small and large {\it main-sequence} galaxies, \Fig{evolution_MS_sizemass_smalllarge}.

In \Fig{evolution_Q_quantities_smalllarge} we further examine the evolution of quenched galaxies, separated by $z=0$ size quartiles, by examining additional quantities along their median evolutionary tracks (similarly to \Fig{evolution_MSQ_quantities} that compares quenched and main-sequence galaxies). The first row, showing $\Delta{\rm SFMS}$ versus the fractional final mass, indicates little separation between large and small quenched galaxies in terms of the fractional mass they reach when they quench: except for the most massive galaxies, there is very little mass growth after quenching (see also \citealp{NelsonD_17a}). The only exception is the $M_{*,z=0}\sim10^{9.4}\Msun$ bin, where the small galaxies show $\approx0.2\dex$ mass loss after they quench. Table \ref{t:numgals} indicates that these galaxies are completely dominated by satellites, hence we may associate this mass loss with stellar stripping. This mass loss can be seen more directly in the second row, which presents fractional mass versus redshift. It also shows that at intermediate and high masses, the smaller galaxies form earlier, at $z\gtrsim1$, while the larger ones quench and stop growing their mass at $z\sim0.5$. At the lowest mass bin of $M_{*,z=0}\sim10^{8.5}\Msun$ the situation is reversed, and it is the larger galaxies that quench earlier. The origin of this difference is elucidated in the bottom two rows of \Fig{evolution_Q_quantities_smalllarge}.

The fourth row in \Fig{evolution_Q_quantities_smalllarge}, showing size versus $\Delta{\rm SFMS}$ color-coded by redshift, demonstrates that for all galaxies with $M_{*,z=0}\sim10^{9-11}\Msun$, almost the entirety of the size growth occurs at $\Delta{\rm SFMS}>-0.5\dex$, namely before quenching, on the main-sequence, for both the large and small galaxies. Hence, galaxies that quench later have more time for size growth, and are larger at $z=0$. In contrast, in the lowest mass bin, the galaxies in the larger quartile experience a significant size growth also after they quench (possibly due to tidal heating, as they are essentially all satellites), hence those quenching earlier are the ones that are larger at $z=0$. The third row, presenting size versus redshift color-coded by $\Delta{\rm SFMS}$, shows that at $M_{*,z=0}\sim10^{9.4}\Msun$ the mass drop discussed above is associated with a drop in size, which is consistent with stripping as the common origin. At $M_{*,z=0}\sim10^{8.5}\Msun$, the small galaxies experience both a mild size drop and an increase, likely as a result of a competition between satellite stripping and heating. At intermediate masses, small galaxies show a slower rate of size evolution with time while their progenitors are on the main-sequence, which combines with the fact that they spend less time on it to result in an overall very small size increase during their main-sequence phase.

\subsection{The Interplay between Size Growth and Quenching}
\label{s:evolution_interplay}
\begin{figure}
\centering
\includegraphics[width=0.475\textwidth]{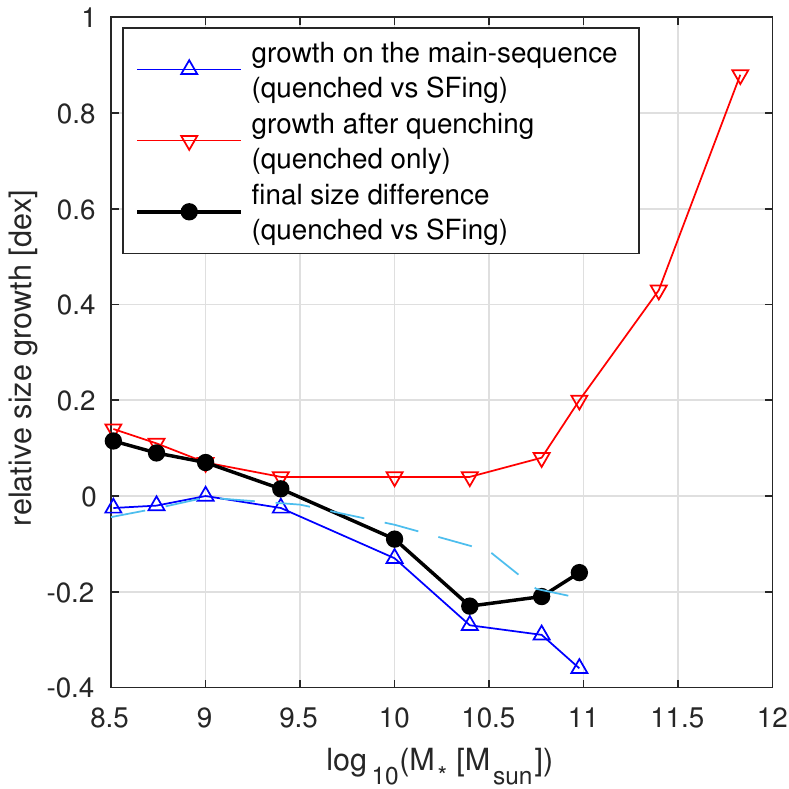}
\caption{The relative size growth between quenched and main-sequence $z=0$ galaxies, separated by the time they spend on the main-sequence (with $\Delta{\rm SFMS}>-0.5\dex$; solid blue), the growth after quenching (with $\Delta{\rm SFMS}<-0.5\dex$, relevant only for quenched galaxies; red), and their sum (black). At low mass, the size growth on the main sequence is very similar whether galaxies end up at $z=0$ as main-sequence or quenched, but a significant gap opens towards higher masses, where $z=0$ quenched galaxies grow significantly less while still on the main-sequence. This difference is partly because at their quenching time, the general population of main-sequence galaxies is smaller than main-sequence galaxies of the same mass at $z=0$ (dashed blue), however in addition they are smaller than that general population (the difference between solid and dashed blue). In the quenched phase, there is very little size growth for intermediate-mass galaxies, but it is significant at both low, and in particular high, masses.}
\vspace{0.3cm}
\label{f:relative_growths}
\end{figure}

\Fig{relative_growths} summarizes several aspects discussed above by showing relative size growth between quenched and main-sequence $z=0$ galaxies, as a function of final mass, and separated by the quenching time. First, the size difference between these two populations at the last time they are still on the main-sequence, which is $z=0$ for main-sequence galaxies and the quenching time (typically around $z\sim1$) for quenched galaxies, is shown in solid blue. Second, the degree of size growth after quenching, which is relevant only for quenched galaxies, is shown in red. Finally, their sum, which by construction represents the size difference between the two populations at $z=0$, is shown in black. The threshold dividing the growth `on' and `below' the main-sequence here is $\Delta{\rm SFMS}\lessgtr-0.5\dex$. At $M_*\lesssim10^9\Msun$, the two galaxy types have very similar sizes at the end of their time on the main-sequence (solid blue), but quenched galaxies have larger $z=0$ sizes (black), because they experience further growth after being quenched (red). This quenched-phase growth diminishes with increasing mass, until at $M_*=10^{9.5}\Msun$ the two galaxy types have equal sizes. Considering the most massive scale, where there are no main-sequence galaxies, quenched galaxies show a degree of quenched-phase size growth that rapidly rises with mass, reaching almost a full order of magnitude of size growth for $M_*\lesssim10^{12}\Msun$ galaxies. This quenched-phase size growth starts to rise around $M_*\sim10^{10.75-11}\Msun$, the scale of the most massive main-sequence galaxies, below which the quenched-phase growth is roughly constant at $\approx0.05\dex$. The main trend at intermediate masses, $M_*\sim10^{9.5-10.75}\Msun$, is that more massive quenched galaxies are progressively smaller at $z=0$ than their main-sequence counterparts (black), due to their diverging degrees of growth while on the main-sequence (blue). 

Quenched galaxies leave the main-sequence at an earlier time than $z=0$, and do so nearly at their final mass (\Fig{evolution_MSQ_quantities}, second row). It is therefore worth asking whether their lesser degree of size growth while on the main-sequence, as quantified in \Fig{relative_growths}, is merely a result of them spending less time on it, or leaving it when the universe was denser, with respect to their $z=0$ main-sequence counterparts. The flat size-mass evolution of quenched galaxies seen in \Fig{evolution_MSQ_sizemass}, as well as their flat size-redshift evolution seen in \Fig{evolution_MSQ_quantities}, suggest otherwise, namely that the main progenitors of $z=0$ quenched galaxies show a particularly slow size evolution even when they are still on the main-sequence. To confirm this explicitly, we begin by introducing an additional curve in \Fig{relative_growths} (dashed blue), which shows the median size of {\it all} main-sequence galaxies selected at the redshift and mass at which the median progenitors of $z=0$ quenched galaxies drop below the main-sequence, relative to the median size of $z=0$ main-sequence galaxies. It is found that at the time and mass at which the progenitors of $z=0$ quenched galaxies undergo quenching, the general population of main-sequence galaxies is indeed smaller than a parallel population at $z=0$. In this sense, the fact that the growth of $z=0$ quenched galaxies is arrested prior to the present epoch indeed contributes to them having a smaller size at $z=0$. However, the size of this general population is not {\it as small} as those progenitors {\it themselves} at the time they quench (solid blue). Hence, a second contribution to the small sizes of $z=0$ quenched galaxies comes from them being smaller than the typical main-sequence galaxy already at the $z>0$ time in which they became quenched.

\begin{figure*}
\centering
\subfigure[$z=0$ main-sequence galaxies]{
          \label{f:mass_z_tracks_MS}
          \includegraphics[width=0.49\textwidth]{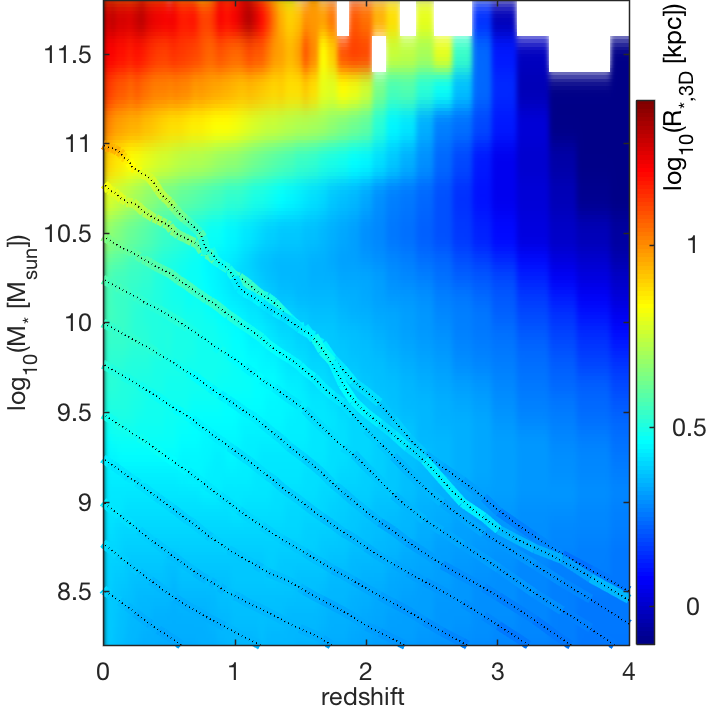}}
\subfigure[$z=0$ quenched galaxies]{
          \label{f:mass_z_tracks_Q}
          \includegraphics[width=0.49\textwidth]{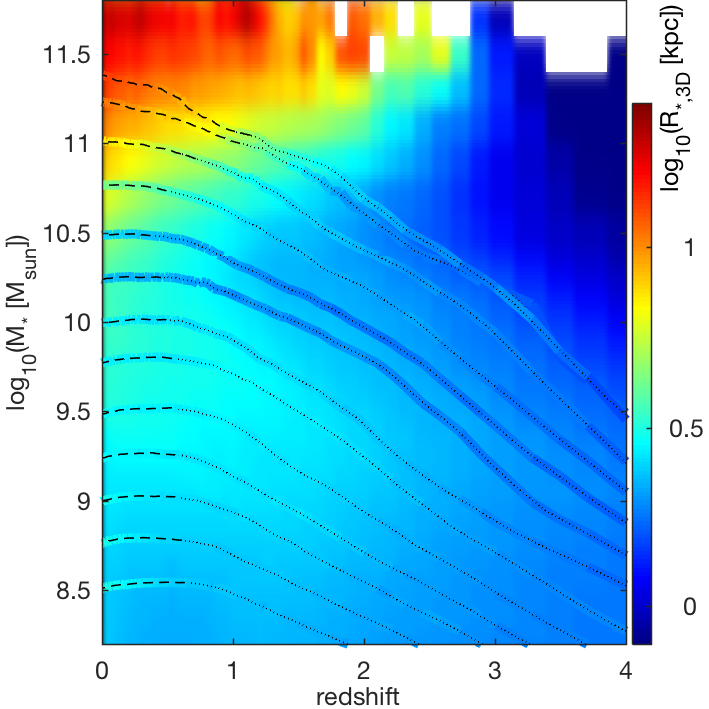}}
\caption{Median evolutionary tracks on the mass-redshift plane of main-sequence (left) and quenched (right) galaxy populations selected with various $z=0$ masses. The colors on each track indicate the median galaxy size at the given redshift, as indicated by the color bar. Each track is also marked with a dotted line where the median $\Delta{\rm SFMS}>-0.5\dex$, and dashed line where the population is already quenched, $\Delta{\rm SFMS}<-0.5\dex$. The background colors show the median size of the {\it full, main-sequence} galaxy population at each mass and redshift. Where the progenitors are on the main-sequence (dotted), the colors of the tracks deviate significantly from the background colors only for tracks with $M_{*,z=0}\sim10^{10.4\pm0.3}\Msun$, both for the progenitors of main-sequence (left) and quenched (right) $z=0$ galaxies. This indicates that the progenitors of galaxies with this mass -- corresponding exactly to where the $z=0$ size difference is maximal -- have biased sizes with respect to the overall main-sequence galaxy population at their corresponding mass and redshift.}
\vspace{0.3cm}
\label{f:mass_z_tracks}
\end{figure*}

\Fig{mass_z_tracks} allows us to delve deeper into the origin of the significant size difference at intermediate masses that develops while the progenitors of both galaxy types are still on the star-formation main-sequence. Evolutionary tracks on the mass-redshift plane (similar to the second row in \Fig{evolution_MSQ_quantities}, only here the mass is not normalized to its final value) are shown for main-sequence (left) and quenched (right) galaxy populations selected with various $z=0$ masses. As discussed for the second row of \Fig{evolution_MSQ_quantities}, main-sequence and quenched galaxies have different shapes of their mass formation histories in that quenched galaxies form their mass earlier. Here, the colors on each track indicate the median galaxy size at that point on the track, and it is also indicated whether the median sSFR still lies within the main-sequence ($\Delta{\rm SFMS}>-0.5\dex$; dotted), or already below it ($\Delta{\rm SFMS}<-0.5\dex$; dashed). In addition, the background colors show the median size of the full (redshift- and mass-dependent) main-sequence galaxy population. The color on most tracks in their main-sequence phase (dotted) lies close to the background color, indicating that the main-sequence main progenitors of $z=0$ galaxies have representative sizes of the full main-sequence populations at the progenitors' mass and redshift. At low masses, the mass and redshift dependences of the sizes of the underlying population are not strong, hence the colors along the progenitor tracks of main-sequence and quenched galaxies are similar too, in spite of the differences that exist in the shape of those tracks.

Important exceptions are, however, the tracks on \Fig{mass_z_tracks} of galaxies with $z=0$ masses around $10^{10.4}\Msun$. We remind the reader that this is the mass scale that shows the maximum difference between the size-mass evolutionary tracks of main-sequence and quenched galaxies (see \Fig{evolution_MSQ_sizemass}). For these galaxies, the progenitor sizes are biased with respect to the overall population of main-sequence galaxies at their respective redshifts and masses. The progenitors of main-sequence galaxies (left) are somewhat biased high in size, and even more so, the progenitors of quenched galaxies (right) are biased low. This means that the widely different size-mass tracks seen in \Fig{evolution_MSQ_sizemass} for this mass range are not solely a result of the different mass formation histories, namely that the progenitors of quenched galaxies at a given mass exist at a higher redshift and are hence naturally smaller. In addition to that, the sizes of the progenitors are biased with respect to the underlying populations in a way that bolsters the differences in the same direction.

To summarize what is seen in \Fig{mass_z_tracks}, the sizes of the progenitors of galaxies selected with the same $z=0$ masses (around $10^{10.4}\Msun$) are smaller for quenched galaxies, at a given progenitor mass, due to the combination of two factors. First, because those progenitors lie at a higher redshift than the progenitors of the same mass of main-sequence galaxies. Second, because the progenitors of quenched galaxies are smaller than the overall population of galaxies at their mass and redshift, while the progenitors of main-sequence galaxies are larger than their corresponding parent population.

\begin{figure}
\centering
\includegraphics[width=0.475\textwidth]{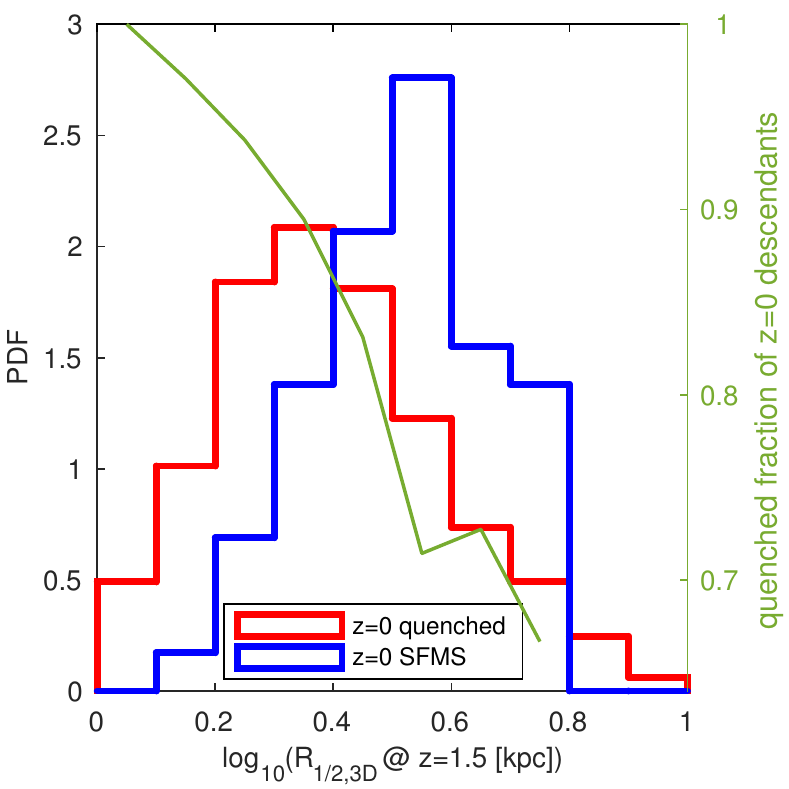}
\caption{Size probability distribution functions for main-sequence $z=1.5$ galaxies in a narrow mass bin around $10^{10}\Msun$, separated according to whether their $z=0$ descendants are quenched (red) or on the main-sequence (blue). A clear correlation exists (with a p-value of $2.8\times10^{-5}$ in the Kolmogorov-Smirnov test) such that smaller $z=1.5$ main-sequence galaxies are more likely to have quenched $z=0$ descendants than their larger counterparts. This is corroborated by the green curve that describes the fraction of quenched $z=0$ descendants as a function of $z=1.5$ galaxy size.}
\vspace{0.3cm}
\label{f:prog_desc_correlation}
\end{figure}

What might be the origin of this phenomenon? The implication for the locations on \Fig{mass_z_tracks} where the colors of the tracks differ from the background color is that galaxies at those particular redshifts and masses `know' about some property of their $z=0$ descendants, and differentiate in size accordingly already at that high redshift. The question opening this paragraph can then be further specified as: what might such properties be, and what might be the physical mechanism for this `knowledge'? To answer this, we select as a concrete example galaxies with $M_*\approx10^{10}\Msun$ at $z=1.5$, which is a regime where both smaller-than-background quenched tracks and larger-than-background main-sequence tracks exist in \Fig{mass_z_tracks}. In other words, the progenitor evolutionary tracks that intersect $z=1.5$ and $M_*\approx10^{10}\Msun$ and are larger than their `background' are those that end up at $z=0$ as main-sequence $M_*\approx10^{10.8}\Msun$ galaxies, while the ones smaller than their `background' end up at $z=0$ as quenched $M_*\approx10^{10.3}\Msun$ galaxies. In \Fig{prog_desc_correlation} we reverse the order in which we have followed the merger trees so far, and examine the sizes of all $z=1.5$, $M_*\approx10^{10}\Msun$ main-sequence galaxies\footnote{Under the condition that they are themselves the main progenitor of a $z=0$ galaxy, rather than merging into a more massive galaxy between $z=1.5$ and $z=0$.} in relation to the star-formation activity of their $z=0$ descendants. \Fig{prog_desc_correlation} shows that those galaxies that end up quenched at $z=0$ tend to have smaller $z=1.5$ sizes than those with main-sequence descendants, by $\approx0.15\dex$. This is a significant difference compared with the size differences seen along the tracks of intermediate-mass galaxies in \Fig{evolution_MSQ_sizemass}. In other words, $z=1.5$ galaxies differentiate in size according to whether they end up quenched or on the main-sequence at $z=0$.

One can speculate that it is the high-redshift size that affects the descendant star-formation activity, rather than some form of `knowledge' that high-redshift galaxies have on their descendant properties that affects their sizes. This may be driven by quenching mechanisms that operate more strongly for galaxies with smaller size, thereby causing the galaxies in which they operate to quench their star-formation and tend to have quenched $z=0$ descendants. A prime suspect for such a mechanism in the TNG100 simulation is black hole feedback; if smaller galaxies (those with higher velocity dispersions at a given mass) harbor more massive black holes (as suggested by observations, and as shown explicitly in our forthcoming work; \citealp{ChangY_17a,KrajnovicD_17a}; Habouzit et al.~in prep.), black hole feedback may be more efficient in galaxies with smaller sizes at suppressing star-formation. This scenario is consistent with the finding that the progenitor tracks in \Fig{mass_z_tracks} that deviate significantly from the background are only those of galaxies with a final mass around $M_*\approx10^{10.5}\Msun$. At smaller masses, black hole feedback is less consequential for star-formation and does not lead to quenching.
At larger masses, black hole feedback becomes effective for essentially {\it all} galaxies, such that small galaxies are not as distinct from the overall population in terms of their future quenching.

\section{Discussion}
\label{s:discussion}
Since it was established that quenched galaxy populations were much more compact at high redshift than they are today \citep{vanDokkumP_07a}, and in particular that the quenched size-mass relation evolves with redshift more rapidly than the star-forming size-mass relation, it has been argued that individual quenched galaxies must undergo significant size growth \citep{FaisstA_17a}, and various physical mechanisms have been proposed as the drivers of this evolution, with a particular emphasis on the role of dry minor mergers \citep{NaabT_09a,HopkinsP_10c}. However, an alternative scenario -- `progenitor bias' -- has also been proposed, where the size evolution of the quenched population can be achieved without the growth of individual quenched galaxies. Instead, given that quenched galaxies at a given mass are observed to be smaller than star-forming ones, the quenched population size growth can be achieved merely by the continuous addition of larger star-forming galaxies as more and more of them quench over time \citep{vanDokkumP_01a,CarolloC_13a}.

The picture that emerges from the study presented here is that individual quenched galaxies do indeed grow in size over cosmic time, but as a population they do so at a rate that does not differ very substantially from individual main-sequence galaxies for most of cosmic time. This conclusion was also reached by \citet{FurlongM_17a} based on the EAGLE simulation. Further, we show that as quenched galaxies grow their mass significantly earlier than main-sequence ones (as indeed also found in the EAGLE simulation; \citealp{ClauwensB_16a}), the respective evolutionary tracks on the size-mass plane differ considerably. At early times, when the progenitors of present-day quenched galaxies are still on the star-formation main-sequence, they grow a large fraction of their mass, but since this occurs over a relatively short fraction of the age of the universe, the size growth is mild, resulting in a shallow evolution on the size-mass plane. This is the pre-quenching phase. After quenching, in contrast, the mass growth slows down significantly (except for the most massive galaxies, which are discussed below), but the size growth continues over time, resulting in steep post-quenching size-mass evolutionary tracks.

\citet{vanDokkumP_15a} used observed data to develop and test a similar scenario for the size evolution of the {\it population} of star-forming galaxies at high redshift. They inferred that as populations, star-forming galaxies at $1.5<z<3$, both `extended' and `compact', evolve on the size-mass plane in `parallel tracks' according to $r\propto M^{0.3}$, and after they quench they experience steeper evolutionary tracks. Our results share with \citet{vanDokkumP_15a} the common framework of two phases of evolution on the size-mass plane, pre- and post-quenching, but differ in the nature of the evolution in the pre-quenching phase. We find that the evolutionary tracks of quenched versus main-sequence galaxies, as well as of quenched (or main-sequence) galaxies of different sizes, are not parallel on the size-mass plane. Instead, the $z=0$ size difference between different populations (either selected by star-formation activity or size itself) tends to decrease as their progenitors are followed backwards in time\footnote{This holds at early enough times, in the main-sequence phase of those progenitors. In the quenched phase, the differences can increase moving backwards in time rather than decrease, as seen in \Figs{evolution_MSQ_sizemass}{evolution_Q_sizemass_smalllarge}.}, until at very high redshift the size-mass tracks almost converge. In other words, quenched galaxies, as well as smaller sized galaxies in general, evolve on shallower tracks than main-sequence galaxies, or larger sized galaxies in general. These quenched/small galaxies also form earlier than their main-sequence/large counterparts. Hence, their progenitors, which start at a similar location on the size-mass plane at high redshift, grow rapidly in mass at early times while not evolving much in size. These two phenomena may be related through a high degree of gas dissipation. Galaxies that grow in mass more gradually tend to grow concurrently in size as well, more in line with a standard `inside-out' growth scenario \citep{MoH_98a}.

\citet{LillyS_16a} propose a possible explanation for the correlation between star-formation activity and size, and negative correlation between stellar age and size, that dismisses the need for a causal relation between the two through a physical mechanism, via the progenitor bias scenario. Our finding that smaller quenched galaxies have earlier formation histories, implying they have quenched earlier in time and with a smaller size, is a manifestation of progenitor bias. We indeed find that what determines the size difference between small and large quenched galaxies (except possibly at masses significantly above $10^{11}\Msun$) is not their post-quenching size growth but their size growth on the main-sequence. Galaxies that grow their mass later also grow to larger sizes while on the main sequence, quench later, and end up at the upper end of the quenched galaxy size distribution. Nevertheless, in TNG100 signatures of the progenitor bias scenario do not tell the full story, as the size evolution of quenched versus main-sequence galaxy populations differ already when they are on the star-formation main-sequence, in that the main-sequence progenitors of quenched galaxies are smaller than other main-sequence galaxies at similar mass and redshift, as shown in \Fig{mass_z_tracks}. This suggests, though indeed does not prove, a causal relation between compactness and quenching, which we postulate, and discuss in more detail in upcoming work, originates from black-hole feedback.

Several differences between the evolutionary tracks of galaxies in TNG100 and in the \citet{LillyS_16a} toy model can give rise to the difference in the role that causality plays (or does not play), as well as differences in the phenomenology. For example, in contrast to \citet{LillyS_16a} the size evolution of the main-sequence population in TNG100 does not scale as $(1+z)^{-1}$ at all masses but the evolution is stronger at higher masses, in qualitative agreement with observations. However, the quantitative agreement is not yet fully borne out in TNG100 either, as the TNG100 $z=2$ size-mass relation is too flat, such that more detailed comparisons of future simulations to similar toy models will still be necessary to distinguish these scenarios with certainty.

Turning to the most massive galaxies in the present day universe, \citet{HillA_17a} used the evolving cumulative number density technique \citep{BehrooziP_13c} to infer past evolutionary tracks of the progenitors of today's extremely massive galaxies of $M_*\approx10^{11.5}\Msun$. The picture they develop has both similarities and differences to the one we find in TNG100. In common with our results, they find that the evolutionary tracks of individual galaxies are steeper on the size-mass plane than the `snapshot' relation at $z=0$, and that the mass and size growth of galaxies of this mass undergoes a transition around $z\sim1.5$ from being dominated by in-situ star-formation to being dominated by dry mergers. In contrast, they however infer significant size evolution of the progenitors of those massive galaxies also during the pre-quenching period, with only a mild steepening of the slope of the size-mass evolutionary track. This difference may stem from a true difference in the evolution of very massive galaxies between the real Universe and TNG100, from limitations of cumulative number density techniques in capturing the correct progenitors (\citealp{TorreyP_14c,WellonsS_17a,TorreyP_17a}, but see \citealp{ClauwensB_17a}), or from other possible observational biases discussed in \citet{HillA_17a}. 

The most massive galaxies we consider quench on average with a stellar mass of $\sim10^{11}\Msun$ (\Fig{evolution_MSQ_quantities}) and from then on grow in size by $\sim1.5\dex$ by $z=0$, when they reach their final mass of $10^{11.8}\Msun$. This is roughly consistent with $r\propto M^2$ as expected from growth by minor dry mergers \citep{NaabT_09a,vanDokkumP_10a,HilzM_13a}. However, we find that while less massive galaxies have a smaller post-quenching size and mass growth than the most massive galaxies, their evolutionary tracks on the size-mass plane are even steeper than $r\propto M^2$. This holds in particular for the smallest quenched galaxies (\Fig{evolution_Q_sizemass_smalllarge}). One physical process that contributes to this steeper slope is stellar mass loss from the old stellar populations that comprise these galaxies. In parallel to the growth of these galaxies by dry mergers, the stellar particles comprising them gradually return mass to the gas phase, thereby limiting the overall stellar mass growth and steepening the size-mass evolutionary track. In addition, this released mass is possibly expelled by black-hole feedback (as these galaxies remain quenched), which may induce a physical expansion of the remaining stellar mass.

At the low-mass end, we find that the slope of the size-mass relation of quenched galaxies is negative at all redshifts. While there is observational evidence for such a trend at $z>0$, it seems that at $z=0$ this is a discrepancy between TNG100 and the observations. As the main-sequence size-mass relation keeps a positive slope also at low masses, at the lowest mass regime we consider quenched galaxies are {\it larger} than main-sequence ones. Considering the evolutionary tracks of these galaxies, this appears to be a result of late-time, post-quenching growth of the low-mass quenched galaxies. Since these galaxies in TNG100 are essentially all ($\gtrsim97\%$) satellites (or splashback), it can be postulated that this size growth is environmentally driven. This satellite fraction is, however, likely too high \citep{PengY_12a}, as the simulation probably misses some low-mass quenched central galaxies \citep{NelsonD_17a,SpringelV_17a}. It is left for future work to determine whether the addition of a population of central low-mass quenched galaxies in the right proportions would change the slope of the quenched low-mass population from negative to positive, as observed. Another possibility is that resolution effects play a role in the size increase of the low-mass quenched satellites \citep{PillepichA_16a}. Future work exploring this regime using simulations with higher resolution will be telling whether this increase is numerical.

Finally, \citet{CharltonP_17a} recently showed evidence for a positive correlation between halo mass and galaxy size, at fixed stellar mass and color, in observations as well as in the EAGLE and Illustris simulations. This is qualitatively consistent with our finding that galaxies with smaller sizes tend to form earlier, as lower-mass halos form earlier than higher-mass ones. However, the effect they find in the simulations is dominated by satellites, and in particular, stripping of satellites, which reduces both their halo masses and galaxy sizes. Our results, however, are similar even if only central galaxies are considered. In addition, the overall correlation they find is too weak to be a significant factor in explaining our results. This can be seen by considering that the size difference between the quartiles in \Fig{evolution_Q_sizemass_smalllarge} is $\sim0.3\dex$, which translates to a $\sim0.15\dex$ difference in halo mass according to \citet{CharltonP_17a}, corresponding to formation time differences of $\Delta z\sim0.05$ (e.g.~\citealp{BorzyszkowskiM_14a}). Such values are not enough to account for the significant differences in formation histories seen in \Fig{evolution_Q_quantities_smalllarge}. In conclusion, then, it seems that the halo mass differences between small and large galaxies found by \citet{CharltonP_17a} are not a significant driver of the results presented here.

The scope of this work is the aggregate growth histories of galaxy populations, but a future natural extension would be to explore the diversity of evolutionary tracks of individual galaxies, and how they combine to the trends presented here. These individual tracks may be complex and `noisy', involving major mergers, compact star-formation configurations triggered by either mergers or internal gravitational instabilities, and star-formation in extended disks, but may also reveal regularities and characteristics that can be categorized into a few main classes (e.g.~\citealp{ZolotovA_15a}). Individual evolutionary tracks may also allow deeper investigations into the nature of the physical processes that shape them.

\section{Summary}
\label{s:summary}
This work explores galaxy sizes in the TNG100 simulation of the IllustrisTNG project \citep{SpringelV_17a,PillepichA_17a,NelsonD_17a,NaimanJ_17a,MarinacciF_17a}, which is treated as an effective model of galaxy formation. This MHD cosmological simulation of a $\sim(110\Mpc)^3$ volume contains many thousands of galaxies and does not make direct assumptions about their sizes, thereby having predictive power in this regard. We find a fair match to observed size relations, and using evolutionary tracks of various galaxy types over time we provide novel scenarios for the size evolution of galaxies.

In the first part of this work, galaxy size-mass relations have been compared to observations separately for main-sequence (namely star-forming) and quenched galaxies, for stellar masses $M_*>10^9\Msun$ and redshifts between $z=0$ and $z=2$. To minimize systematics with respect to the observations, projected circularized r-band half-light sizes, which differ from the `intrinsic' three-dimensional half-mass sizes, have been employed for the comparison. As observed, we recover in the simulation that at fixed stellar mass, quenched galaxies are smaller than main-sequence ones and galaxy sizes become larger with cosmic time. The slope of the size-mass relation steepens with increasing galaxy mass. In particular, the simulated main-sequence galaxies have an approximately constant slope of the size-mass relation, but this slope is shallower than the observed one, and more so towards higher redshifts. The size-mass relation of quenched galaxies is very different, having a negative slope below $M_*\sim10^{9.5}\Msun$, and a steep positive slope thereafter, in general agreement with observations. Across the explored parameter space, the deviations between the simulation and observations do not exceed $0.25\dex$ and are mostly significantly smaller than that.

This confirmation is used as a solid starting point for a study of how galaxy populations, selected at $z=0$ as either main-sequence or quenched, at different masses and sizes, evolve (as an ensemble) over time, with particular emphasis on their size evolution and its relation to the evolution of mass and star-formation activity. The main results from this analysis are the following.

\begin{itemize}
\item Galaxy populations with low $z=0$ mass ($M_*\lesssim10^{9.5}\Msun$) evolve on very similar tracks on the size-mass plane as long as they are on the main-sequence, whether they are still on the main-sequence at $z=0$ or have quenched by then. The $z=0$ quenched galaxies typically form earlier, but this hardly translates to a different size at a given progenitor mass, since the redshift evolution of the overall size-mass relation is very weak at low masses. These $z=0$ quenched galaxies quench around $z\sim1$ and experience very little mass growth thereafter. They do, however, experience size growth of $\lesssim0.2\dex$, making them larger at $z=0$ than their main-sequence counterparts. Low-mass quenched galaxies in TNG100 are almost entirely all satellites, hinting at the possible role of tidal effects in their late-time size growth.
\item Galaxies at $z=0$ of intermediate masses ($M_*\sim10^{10.5}\Msun$) show the largest size difference between quenched and main-sequence populations. This is in part because quenched galaxies have earlier formation histories, such that their progenitors at a given progenitor mass exist at higher redshift than those of main-sequence galaxies, and are hence naturally smaller. Consequently, they also reach their final mass earlier in time, and therefore reach it with a smaller size. This scenario alone is one of `parallel tracks' on the size-mass plane, with earlier-forming galaxies (that are quenched by $z=0$) evolving on a `smaller', parallel track to later-forming ones. However, this is only one part of the picture, and indeed the size-mass evolutionary tracks are in fact not parallel.
\item Instead, quenched and main-sequence galaxies at intermediate mass show the largest deviations also in their size evolution histories between these two types. While galaxy populations that are on the main-sequence all the way to $z=0$ grow with a roughly constant slope on the size-mass plane, galaxies that end up quenched at $z=0$ show two distinct phases in their size evolution, switching from one to the other around their time of quenching. Even before they quench, while they are still on the main-sequence, their size evolution differs from that of galaxies that will remain on the main-sequence down to $z=0$. In this phase, their main progenitors grow by orders of magnitude in mass but with a nearly flat size-mass slope, and they have smaller sizes than the general population of main-sequence galaxies of their (evolving) mass and redshift. Hence, quenched galaxies have smaller $z=0$ size not only because at fixed progenitor mass they reside at higher redshift, but also because they are smaller than their parent main-sequence populations already at those high redshifts. Indeed, the reverse is found to hold as well, namely the $z=0$ descendants of smaller $z>0$ galaxies are preferentially quenched. This hints either to something in their formation histories that in addition to making them early-formers also sets their sizes to be small already at very early times, or at a causal relationship between small size and quenching, such that small galaxies have their star-formation arrested. A prime candidate for the driver of the latter scenario is AGN feedback, as will be explored more directly in upcoming work.
\item After their quenching, galaxy populations of intermediate mass and above ($M_*\gtrsim10^{10.5}\Msun$) show a phase of size growth that is characterized by a steep size evolution as a function of the growing stellar mass. Intermediate-mass galaxies grow little in mass in the quenched phase, hence also their size growth in this phase is not large. Higher mass galaxies progressively experience more mass growth in the quenched phase, and accordingly also more size growth. Interestingly, lower-mass quenched galaxies, and those that are smaller at $z=0$, are actually those that show the steepest size-mass evolutionary tracks in their quenched phase. For $M_*\sim10^{10-11}\Msun$ galaxies, the dominant factor in their being smaller is the degree of size growth they experienced while they were still main-sequence galaxies. The size growth of the most massive galaxies, $M_*\gtrsim10^{11}\Msun$, however, occurs mostly during their quenched phase.
\end{itemize}

\section*{Acknowledgements}
SG thanks Greg Bryan, Avishai Dekel, Andrey Kravtsov, and Rachel Somerville, for useful discussions. We are grateful to Vicente Rodriguez-Gomez for sharing the codes to generate the merger trees for the simulation. The Flatiron Institute is supported by the Simons Foundation. SG and PT acknowledge support from NASA through Hubble Fellowship grants HST-HF2-51341.001-A and HST-HF2-51384.001-A, respectively, awarded by the STScI, which is operated by the Association of Universities for Research in Astronomy, Inc., for NASA, under contract NAS5-26555. VS, RW, and RP acknowledge support through the European Research Council under the ERC-StG grant EXAGAL-308037 and would like to thank the Klaus Tschira Foundation. VS also acknowledges support through subproject EXAMAG of the Priority Programme 1648 `Software for Exascale Computing' of the German Science Foundation. RW acknowledges support by the IMPRS for Astronomy and Cosmic Physics at the University of Heidelberg. LH acknowledges support from NASA grant NNX12AC67G and NSF grant AST-1312095. JPN acknowledges support of NSF AARF award AST-1402480. MV acknowledges support through an MIT RSC award, the support of the Alfred P.~Sloan Foundation, and support by NASA ATP grant NNX17AG29G. TNG100 was run on the HazelHen Cray XC40-system at the High Performance Computing Center Stuttgart as part of project GCS-ILLU of the Gauss Centre for Supercomputing (GCS). Ancillary and test runs of the IllustrisTNG project were also run on the Stampede supercomputer at TACC/XSEDE (allocation AST140063), at the Hydra and Draco supercomputers at the Max Planck Computing and Data Facility, and on the MIT/Harvard computing facilities supported by FAS and MIT MKI.

\appendix

\section{The Star-Formation Main-Sequence}
\label{s:main_sequence}
\begin{figure}
\centering
\includegraphics[width=0.475\textwidth]{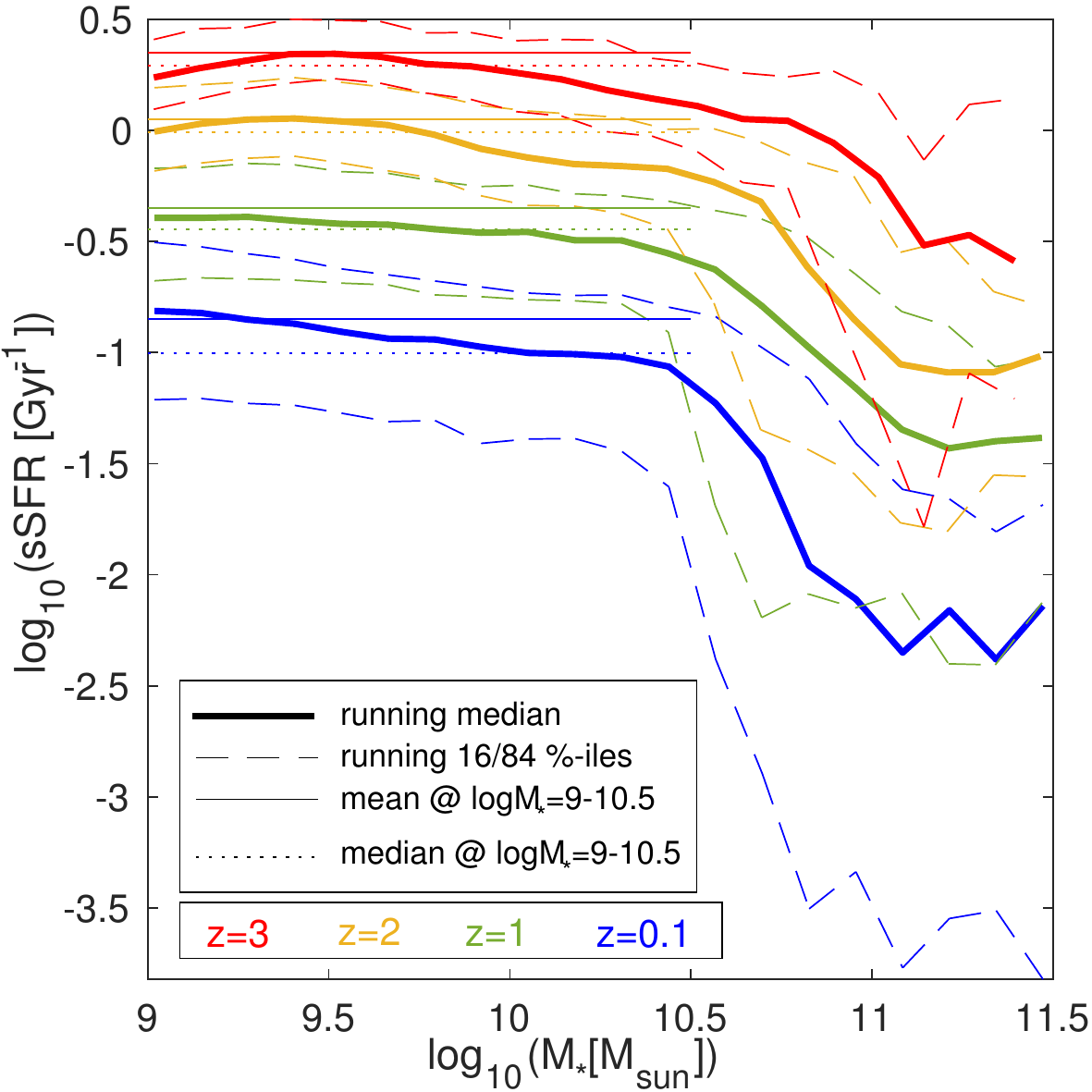}
\caption{Statistics of the sSFR as a function of stellar mass for $z=0.1,1,2,3$, which are used to define `main-sequence' and `quenched' galaxies in Section \ref{s:comparison}. For the calculation of the running median (thick solid) and $16^{\rm th}-84^{\rm th}$ percentiles (dashed), only galaxies with a non-zero instantaneous SFR are included. When galaxies with zero SFR are included too, the $16^{\rm th}$ percentile equals zero at large masses and low redshift.} 
\vspace{0.3cm}
\label{f:main_sequence}
\end{figure}

\Fig{main_sequence} displays several statistics of the sSFR as a function of stellar mass for $z=0.1,1,2,3$. At $M_*\lesssim10^{10.5}\Msun$ the running median (thick solid) and spread around it (dashed) represent the main sequence of star-forming galaxies, while at higher masses the various percentiles of the distribution decrease significantly due to the increasing proportion of quenched galaxies (see \citet{NelsonD_17a} for a related detailed account of the color-mass diagram in TNG100). The agreement with observations is good at $z=0.1$ in both normalization and slope. Note that the average slope at $M_*\lesssim10^{10.5}\Msun$ is $\approx(-0.1)-(-0.2)$ at all redshifts, in general agreement with observations (e.g.~\citealp{KurczynskiP_16a}). This represents a mild improvement upon the original Illustris simulation, where the slope was at most $-0.08$ \citep{GenelS_14a}. However, at $z>0$ the normalization is lower than what most observations indicate, a tension that is in common to all large-volume cosmological hydrodynamical simulations (e.g.~\citealp{FurlongM_15a,DaveR_17a}).

Two horizontal lines for each redshift illustrate possible simple definitions for the `ridge' of the main-sequence. {\bf (i)} The mean of the sSFR for all galaxies with $10^{9}\Msun<M_*<10^{10.5}\Msun$ (thin solid), which is the fiducial definition we use in Section \ref{s:comparison} to separate galaxies in `main-sequence' and `quenched'. {\bf (ii)} The median sSFR of all galaxies with $10^{9}\Msun<M_*<10^{10.5}\Msun$, including those that have SFR$=0$ (dotted). This results in lower values, which if adopted do not affect however the relations presented in Section \ref{s:comparison} by more than a few percent.

\section{More Detailed Evolutionary Tracks of Main-Sequence versus Quenched Galaxies}
\label{s:evolution_MSQ_appendix}

\begin{figure*}
\centering
\includegraphics[width=1.0\textwidth]{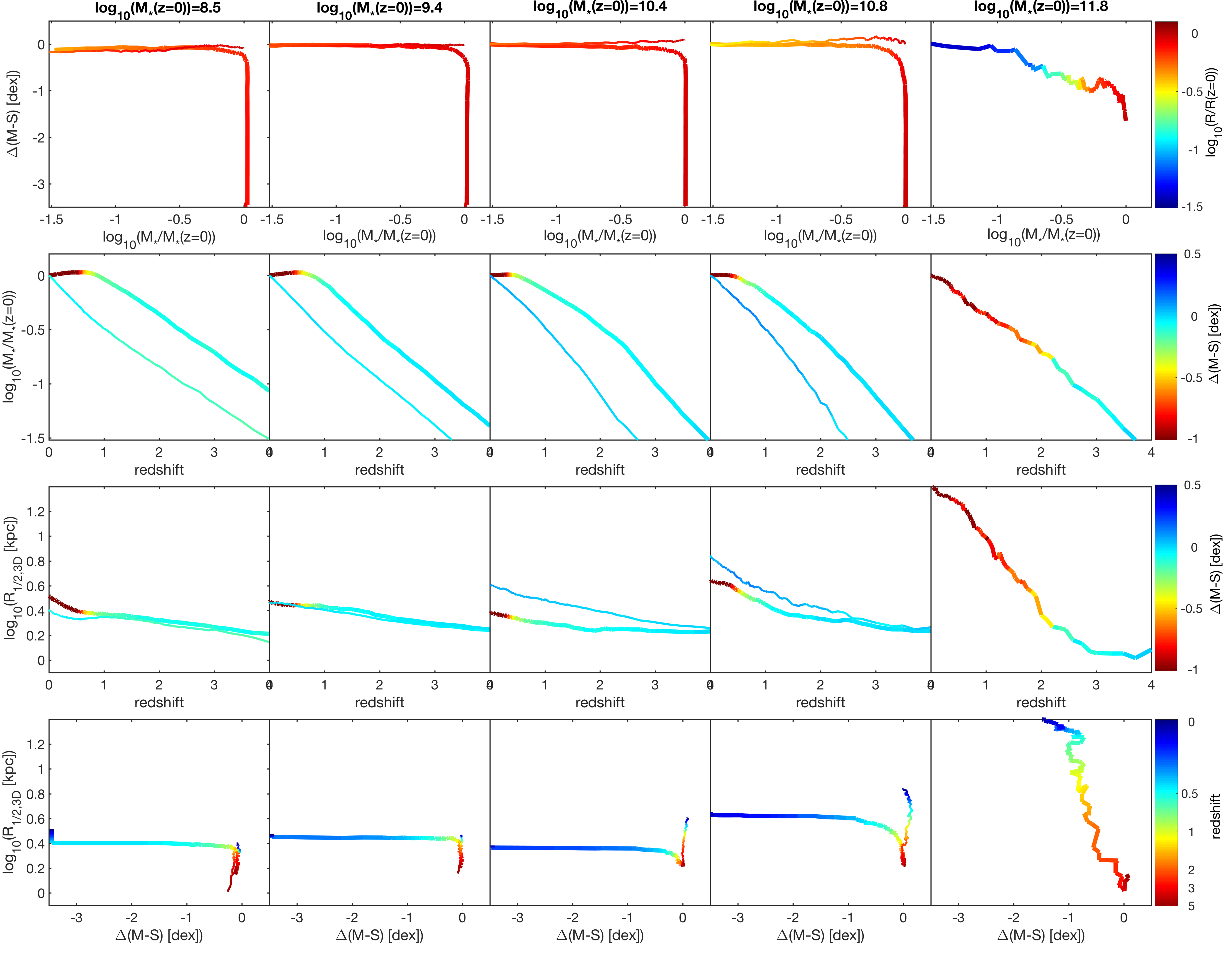}
\caption{Similar to \Fig{evolution_MSQ_sizemass}, namely main-progenitor evolutionary tracks of galaxy populations selected with increasingly larger $z=0$ masses (left to right), where in each panel (except the right-most ones) there appears a track for quenched galaxies (thick curve) and a track for main-sequence galaxies (thin curve). Here, each row presents a different combination of quantities on the vertical and horizontal axes as well as for the color coding that are the same as in \Fig{evolution_Q_quantities_smalllarge} (where they are shown for quenched galaxies separated by their final size). Here it is seen that quenched galaxies, compared to main-sequence ones, reach their final size at a larger fraction of their final mass (first row), form their mass earlier in time (second row), and have flatter size growth rates (except at the highest masses, where no main-sequence galaxies exist; third row). The degree to which quenched galaxies grow their mass before versus after they become quenched is strongly mass dependent (fourth row and \Fig{relative_growths}).}
\vspace{0.3cm}
\label{f:evolution_MSQ_quantities}
\end{figure*}

To provide more quantitative and explicit details on the evolutionary tracks of main-sequence versus quenched galaxies that are presented in \Fig{evolution_MSQ_sizemass}, \Fig{evolution_MSQ_quantities} presents similar tracks, for the same galaxy samples, but with various combinations of quantities on the horizontal, vertical, and `color' axes. The first row shows $\Delta{\rm SFMS}$ versus the fractional final mass, namely the progenitor mass normalized by the $z=0$ mass, color-coded by fractional final size. The tracks of main-sequence galaxies appear as horizontal lines, showing directly what could be read from the colors in the bottom row of \Fig{evolution_MSQ_sizemass}, namely that their progenitors always lie, in the median, on the main-sequence. The tracks of quenched galaxies show that with the exception of the most massive ones, quenched systems have on average very little mass growth after they have been quenched, as their $\Delta{\rm SFMS}$ values only drop when they are very close to their final mass. Also, with the exception of the most massive galaxies, quenched galaxies are very close to their final {\it size} when their $\Delta{\rm SFMS}$ drops, i.e.~they quench. In other words, most of both the mass and size growth of quenched galaxies with $M_{*,z=0}\lesssim10^{11}\Msun$ occurs while they are still on the main-sequence. The lack of mass growth after quenching can also be seen in the second row, which shows fractional final mass versus redshift, color-coded by $\Delta{\rm SFMS}$. This row also shows explicitly that quenched galaxies have earlier formation histories: their progenitors are consistently more massive at a given time than those of their main-sequence counterparts, and at all $M_{*,z=0}$ (with the exception of the most massive ones) they tend to quench and stop growing in mass around $z\sim1$. Main-sequence galaxies, in contrast, grow in mass continuously, and at $z=1$ have only reached $\sim1/3$ of their final mass.

The third row in \Fig{evolution_MSQ_quantities}, presenting size versus redshift color-coded by $\Delta{\rm SFMS}$, shows that at $M_*\lesssim10^{10}\Msun$ the sizes of quenched and main-sequence galaxies evolve in redshift very similarly while they are on the main-sequence. At higher masses, however, the sizes of quenched galaxies evolve {\it slower} with redshift, with respect to main-sequence galaxies, in particular at high redshifts, {\it but even after they quench}. Namely, the steep size-mass evolutionary tracks of intermediate-mass quenched galaxies after their quenching time, which are seen in \Fig{evolution_MSQ_sizemass}, do not allow them to close the size gap from main-sequence galaxies, as their size evolutionary tracks as a function of redshift are in fact not steeper at all. Only at $M_{*,z=0}\gtrsim10^{11}\Msun$, where no main-sequence galaxies exist, the progenitors of quenched galaxies experience a rapid size growth, which occurs after they have been quenched. Earlier than that, their size hardly evolves with redshift. Finally, in the fourth row of \Fig{evolution_MSQ_quantities}, which shows size versus $\Delta{\rm SFMS}$ color-coded by redshift, main-sequence galaxies appear as vertical lines, indicating a gradual size growth on the main-sequence, and quenched galaxies are seen to experience some size growth while on the main-sequence, too. In fact, it can be seen directly that at $M_{*,z=0}\lesssim10^{11}\Msun$, most of the size growth of quenched galaxies occurs on the main sequence (where $\Delta{\rm SFMS}\gtrsim-0.5\dex$), rather than in their quenched phase. Only more massive galaxies experience most of their size growth below the main-sequence at $\Delta{\rm SFMS}<-0.5\dex$.

\section{Evolutionary tracks of Small and Large Main-Sequence Galaxies}
\label{s:evolution_MS_smalllarge}

\begin{figure*}
\centering
\includegraphics[width=1.0\textwidth]{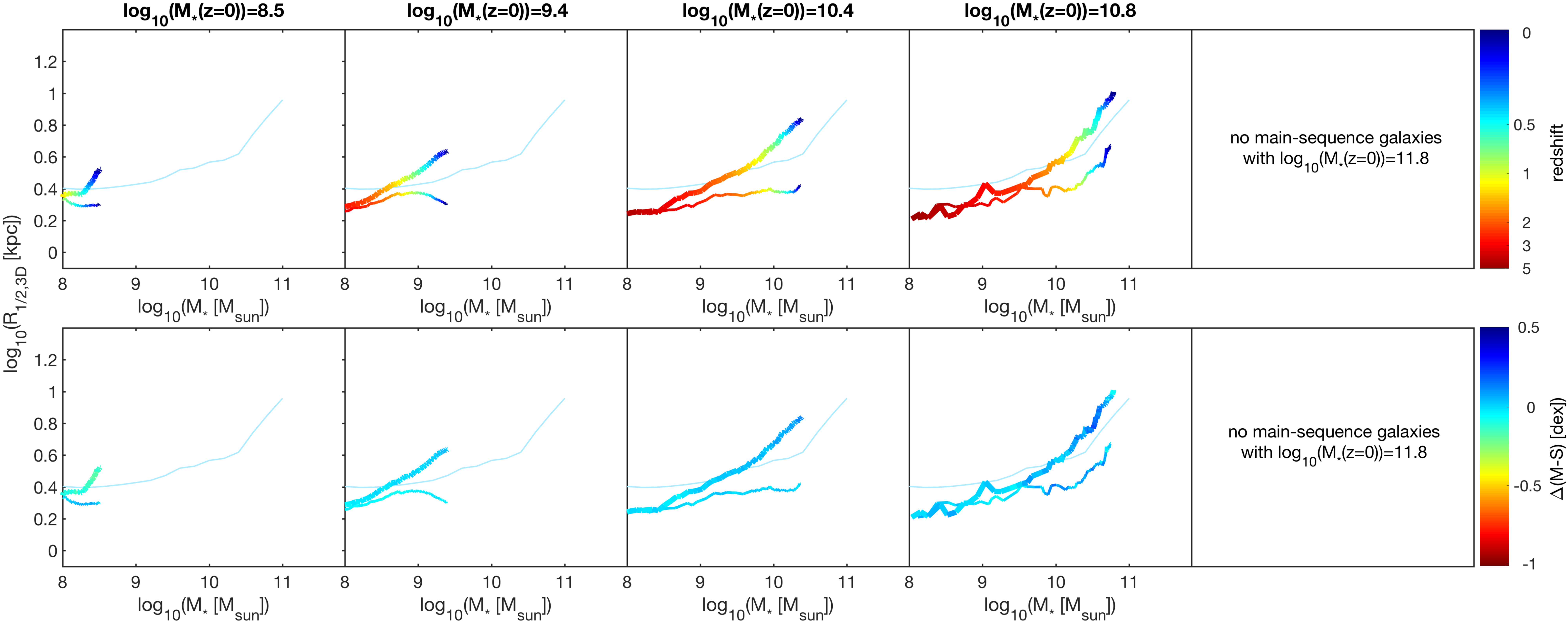}
\caption{Similar to \Fig{evolution_Q_sizemass_smalllarge}, only for main-sequence galaxies rather than quenched ones.}
\vspace{0.3cm}
\label{f:evolution_MS_sizemass_smalllarge}
\end{figure*}

\Fig{evolution_MS_sizemass_smalllarge} is similar to \Fig{evolution_Q_sizemass_smalllarge}, namely with selections in the upper and lower quartiles in terms of $z=0$ size, but for main-sequence galaxies. It shows that the median evolution of these two selections of main-sequence galaxies separates already at high redshift, and they both lie consistently on the main-sequence, but evolve with different slopes of the size-mass plane.

\label{lastpage}

\end{document}